\documentclass[11pt]{article}

\usepackage{amssymb,amsmath,amsthm,amsfonts,graphicx}

\usepackage[T1]{fontenc}
\usepackage[tt=false, type1=true]{libertine}
\usepackage[varqu]{zi4}
\usepackage[libertine]{newtxmath}

\usepackage[margin=1in]{geometry}
\usepackage{enumitem}
\setlist{nosep,topsep=0pt,leftmargin=*}

\usepackage{hyperref}
\hypersetup{
    colorlinks,
    allcolors=teal
}

\newenvironment{myproof}[1][Proof.]{%
    \begin{proof}[#1]%
}{%
    \end{proof}%
}

\usepackage[sortcites,sorting=nyt,style=alphabetic]{biblatex}
\newtheorem{theorem}{Theorem}[section]

\newtheorem{lemma}[theorem]{Lemma}

\newtheorem{remark}[theorem]{Remark}

\theoremstyle{definition}

\newtheorem{definition}{Definition}[section]

\newcommand{\citet}[2][]{\textcite[#1]{#2}}
\newcommand{\citep}[2][]{\cite[#1]{#2}}

\title{Liquid Welfare Guarantees for No-Regret Learning in Sequential Budgeted Auctions}

\author{
    Giannis Fikioris
    \thanks{Supported in part by AFOSR grant FA9550-19-1-0183 and FA9550-23-1-0068, the Department of Defense (DoD) through the National Defense Science \& Engineering Graduate (NDSEG) Fellowship Program, and the Onassis Foundation -- Scholarship ID: F ZS 068-1/2022-2023.}\\
    Cornell University\\
    \texttt{gfikioris@cs.cornell.edu}
    \and
    \'Eva Tardos
    \thanks{Supported in part by NSF grant CCF-1408673, AFOSR grants FA9550-19-1-0183, FA9550-23-1-0410, and FA9550-23-1-0068.}\\
    Cornell University\\
    \texttt{eva.tardos@cornell.edu}
}

\date{Accepted in \href{https://dl.acm.org/doi/10.1145/3580507.3597772}{\textcolor{black}{EC'23}} and \href{https://pubsonline.informs.org/doi/full/10.1287/moor.2023.0274}{\textcolor{black}{Mathematics of Operations Research}}}

\usepackage{nicefrac,xcolor,booktabs}
\usepackage[bb=dsserif]{mathalpha}
\usepackage[capitalize]{cleveref}

\let\g\gamma
\let\e\epsilon
\let\d\delta
\let\l\lambda
\let\sub\subseteq

\newcommand{\lw}{\mathtt{LW}}
\newcommand{\reg}{\mathtt{Reg}}
\newcommand{\optfa}{\mathtt{OPT}_\mathtt{FA}}
\newcommand{\rew}{\mathtt{REW}}
\newcommand{\N}{\mathbb{N}}
\newcommand{\R}{\mathbb{R}}
\newcommand{\calT}{\mathcal{T}}
\newcommand{\calS}{\mathcal{S}}
\newcommand{\calO}{\mathcal{O}}

\newcommand{\Ex}[2][]{\mathbb{E}_{#1}\left[#2\right]}
\renewcommand{\Pr}[1]{\mathbb{P}\left[#1\right]}
\newcommand{\One}[1]{\mathbb{1}\left[#1\right]}
\newcommand{\marg}[2]{\left(#1 \,\big|\, #2\right)}

\usepackage[showdeletions]{color-edits} 
\definecolor{@gray}{HTML}{edc3c5}
\addauthor[Eva]{et}{orange}
\addauthor[Giannis]{gf}{teal}

\begin{document}

\maketitle{}

\begin{abstract}
    We study the liquid welfare in sequential first-price auctions with budget-limited buyers. We focus on first-price auctions, which are increasingly commonly used in many settings, and consider liquid welfare, a natural and well-studied generalization of social welfare for the case of budget-constrained buyers. We use a behavioral model for the buyers, assuming a learning style guarantee: the resulting utility of each buyer is within a $\gamma$ factor (where $\gamma \ge 1$) of the utility achievable by shading her value with the same factor at each iteration. Under this assumption, we show a $\gamma + 1/2 + O(1/\gamma)$ price of anarchy for liquid welfare assuming buyers have additive valuations. This positive result is in stark contrast to sequential second-price auctions, where even with $\gamma=1$, the resulting liquid welfare can be arbitrarily smaller than the maximum liquid welfare, even though the latter can be achieved by a constant shading multiplier. We prove a lower bound of $\gamma$ on the liquid welfare loss under the above assumption in first-price auctions, making our bound asymptotically tight. For the case when $\gamma = 1$ our theorem implies a price of anarchy upper bound that is about $2.41$; we show a lower bound of $2$ for that case.

We also give a learning algorithm that the players can use to achieve the guarantee needed for our liquid welfare result. Our algorithm achieves utility within a $\gamma = O(1)$ factor of the optimal utility even when a buyer's values and the bids of the other buyers are chosen adversarially, assuming the buyer's budget grows linearly with time. The competitiveness guarantee of the learning algorithm deteriorates somewhat as the budget grows slower than linearly with time. 

Finally, we extend our liquid welfare results for the case where buyers have submodular valuations over the set of items they win across iterations with a slightly worse price of anarchy bound of $\gamma + 1 + O(1/\gamma)$ compared to the guarantee for the additive case.    
\end{abstract}

\section{Introduction}

In 2022, over 90\% of all digital display ad dollars were transacted programmatically and accounted for over \$123 billion in the United States \cite{yuen_2022}.
This is usually done through auction platforms where the advertisers subsequently submit bids to place their ads at slots, as the latter become available.
This is done in many platforms such as Google’s DoubleClick, more centralized platforms such as Facebook Exchange and Twitter Ads, as well as sponsored search platforms such as Google’s AdWords and Microsoft’s Bing Ads.
Many of these platforms have started transitioning towards a first-price auction.
For example, in 2021 Google's AdSense announced plans that they will start using a first-price auction format instead of a second-price one \citep{wong_2021} due to the simplicity and transparency of first-price (see also \citep{alcobendas2021adjustment,despotakis2021first,choi2020online}).
In these platforms, advertisers participate in thousands of auctions per day, and their bidding must take into account the limited budget that they have available.
Because of this complexity, advertisers usually deploy automated bidding algorithms that, given the player's budget and preferences, bid in these sequential auctions.
A common strategy in many automated bidding algorithms is to use a \textit{shading multiplier} \citep{DBLP:journals/ior/ConitzerKSM22,DBLP:journals/mansci/ConitzerKPSMSW22,DBLP:journals/mansci/BalseiroG19,DBLP:journals/mansci/BalseiroKK23}: from a single player's perspective, every round the algorithm bids by shading the true value by some multiplicative factor, ensuring winning items of only high value relative to the price.
Formally, using $\l \ge 0$ as a shading multiplier means bidding $\l$ times the value, and the best-fixed shading multiplier in retrospect is the shading multiplier that maximizes a player's utility.

While each player aims to maximize her utility, one of the most common goals of the auction platform is to maximize welfare.
When players are budget-limited we cannot expect low-budget players to achieve high welfare.
Instead, a standard welfare metric used in this case is \textit{liquid welfare} \citep{DBLP:conf/icalp/DobzinskiL14}, the generalization of social welfare, giving credit to welfare achieved by players only up to their budget.
Formally, a player's contribution to the liquid welfare is the minimum of her accumulated value and her budget.
This work studies liquid welfare guarantees when budget-limited players use online learning algorithms to learn a good shading multiplier as they participate in sequential auctions.

This question was studied by \citet{DBLP:journals/corr/GaitondeLLLS22}, who show that if every player uses the online learning algorithm of \cite{DBLP:journals/mansci/BalseiroG19}, then, in both first and second price auctions, the resulting liquid welfare is half the optimal one.
However, their result requires strong assumptions: the players' values need to be identical and independent across rounds, and most importantly, players need to use a specific algorithm.

In this paper, we aim to relax both of the assumptions used by \cite{DBLP:journals/corr/GaitondeLLLS22}.
First, we do not make any assumptions about the values of the players, except that they are upper bounded by $1$.
Second, we replace the assumption that the players use a specific algorithm with a more general behavioral assumption.
An appealing and weaker behavioral assumption is to assume that players have small regret or (more generally) small competitive ratio in the utility they achieve, without requiring that they use a particular algorithm.
\cite{DBLP:journals/corr/GaitondeLLLS22} prove that in sequential budgeted second-price auctions, even when players have no-regret, the resulting liquid welfare can be arbitrarily bad compared to the optimal one (see \cref{thm:second_price}).

In contrast, we focus on sequential budgeted first-price auctions, for which we show that no-regret, and even bounded competitive ratio, do imply liquid welfare guarantees.
We assume that players have bounded competitive ratio with respect to the best-fixed shading multiplier in retrospect.
The optimal shading multiplier is a weak benchmark, e.g., it is much weaker than the best sequence of bids, which implies that our assumption that the players have bounded competitive ratio with respect to this benchmark is easier to satisfy compared to other benchmarks, making it a more appealing behavioral assumption.
We summarize our main results next.

\begin{itemize}
    \item Our first and main result is in \cref{sec:guar}, that in sequential budgeted first-price auctions if every player is guaranteed a competitive ratio of $\g$ compared to using the best-fixed shading multiplier (while budget lasts), then the liquid welfare is within a factor of $\g + 1/2 + O(\g^{-1})$ of the optimal one (\cref{thm:guar}).
    When $\g = 1$, the same theorem guarantees a factor of about $2.41$, which is close to the bound of $2$ of \cite{DBLP:journals/corr/GaitondeLLLS22} and \cite{DBLP:journals/mansci/BalseiroKK23}, but in a less constrained setting: players can employ any competitive strategy (not just a specific algorithm), they are not necessarily in equilibrium, and their values can be picked arbitrarily.
    
    The techniques we use to prove these theorems differ from the ones used by \cite{DBLP:journals/corr/GaitondeLLLS22}, where they assume i.i.d. values across rounds and that players use a specific algorithm.
    Their proofs focus on which rounds a player bids less than her value, how long this behavior lasts, and how ``efficient'' her spending in each round is (specifically they prove their algorithm does not overspend while yielding a high utility each round on expectation).
    We cannot rely on such properties in our setting since values might be adversarially picked and players can have arbitrary behavior in each round (we only require that their resulting utility is competitive).
    In contrast, we compare the utility achieved by each player to the utility possible for a well-chosen shading multiplier $\l\in (0,1)$.
    The competitive ratio assumption guarantees that a player's utility is comparable to what would have been achievable with such a fixed shading.
    Next, we distinguish two cases to prove that the latter is high.
    If by using this shading multiplier $\l$ a player would not run out of budget, then her utility is high using ideas similar to the classic price of anarchy work (e.g., \cite{DBLP:conf/stoc/SyrgkanisT13}).
    In contrast, if that multiplier would have made the player constrained by her budget then we use a very different argument to show high utility: the resulting value is $1/\l$ times the payment, which is high since the budget is almost depleted.
    
    \item In \cref{sec:imp}, we prove a lower bound that almost matches our upper bound: if every player has competitive ratio $\g$ or worse then the resulting liquid welfare can be $\max\{\g, 2\}$ times smaller than the optimal one, for any $\g \ge 1$ (\cref{thm:imp:2,thm:imp:gamma}).
    Both of these theorems hold even when the players have bounded competitive ratio with respect to the best sequence of bids.
    This shows that one cannot guarantee liquid welfare much higher than what our upper bounds imply, even with a stronger behavioral assumption.
   
    \item In \cref{sec:algo}, we offer a bidding algorithm for sequential first-price auctions with budgets.
    When players are budget-limited and the setting (in our case the players' values) is picked adversarially, no-regret guarantees are not achievable (proven for second-price by \cite{DBLP:journals/mansci/BalseiroG19} whose counter-example works in our setting as well).
    In contrast, work focuses on guarantees about the competitive ratio $\g \ge 1$.
    We prove that when the total number of rounds is $T$, a player with budget $B$ can achieve a competitive ratio of $T/B$ with high probability assuming $B = T^{1/2+\Omega(1)}$, even when her values and the other players' bids are adversarially picked (\cref{thm:algo}).
    We note that when the budget grows linearly with time, i.e., $B = \Omega(T)$, this is the same competitive ratio that \cite{DBLP:journals/mansci/BalseiroG19} achieve in expectation for second-price auctions, which they prove is tight in that case.
    
    To achieve this result we reduce sequential first-price auctions to the Bandits with Knapsack (BwK) setting \citep{DBLP:journals/jacm/BadanidiyuruKS18}.
    We overcome a few differences between the two settings for this reduction.
    Our action space is the set of possible bids, which is continuous.
    The known results for BwK for continuous action spaces assume that, as a function of the action, the rewards are concave and the consumption of the resource is convex, which is not true in our case: our rewards (player's utility) and consumptions (player's payment) are neither concave nor convex.
    Additionally, in BwK, an action that would lead to over-consumption of the resource ends the game, while in our setting the bid is adjusted to the remaining budget, and the game continues.
   
    \item Finally, in \cref{sec:subm}, we extend our first result to the case when players have submodular valuations instead of additive ones.
    In this case, we prove that if every player achieves a competitive ratio of $\g$, then the liquid welfare is within a factor of $(2 + \g + \sqrt{\g^2 + 4})/2$ of the optimal one (\cref{thm:subm}), a bound that is a little worse than our bound of $\g + O(1)$ for the linear case.
\end{itemize}
\section{Related Work}
\label{sec:related}

Assuming that players' behavior is based on using a shading multiplier to shade one's value has recently received a lot of attention.
\citet{DBLP:journals/mansci/BalseiroG19} propose an adaptive pacing mechanism for sequential second-price auctions that shades one's bid based on the total budget and payment up to that round.
From a single player's perspective, they prove that even if values and prices are picked adversarially, the player can guarantee a competitive ratio of $\max\{1, \frac{T}{B}\}$ compared to her maximum utility in retrospect, assuming the maximum value she can have each round is $1$, $B$ is her budget, and $T$ is the number of rounds; they also prove that this bound is tight.
For the case where values and prices are picked from the same distribution every round, they prove a competitive ratio of $1$ and $O(\sqrt T)$ regret.
They also prove that each player achieves a competitive ratio of $1$ when every player is using their pacing algorithm and their values are sampled from distributions that satisfy certain properties.
In repeated second-price auctions using a fixed shading multiplier to bid can yield the optimal utility when the player is not budget-limited (which is useful in adversarial budgeted settings) or when the player's values are continuously distributed.
While this is not true for first-price, the simplicity of considering shading multipliers to manage one's budget motivates using this bidding strategy in this case also, see for example the works of \citet{DBLP:journals/mansci/ConitzerKPSMSW22,DBLP:journals/mansci/BalseiroKK23,DBLP:journals/corr/abs-2203-16816}.
Similar work for bidding in second price auctions has been done by \citet{DBLP:journals/corr/abs-2207-04690} where they consider mechanisms that use budget throttling instead of pacing multipliers with guarantees similar to \citet{DBLP:journals/mansci/BalseiroG19}.

\citet{DBLP:journals/corr/GaitondeLLLS22} focus on guarantees of the pacing algorithm of \citet{DBLP:journals/mansci/BalseiroG19} in sequential budgeted auctions.
They prove that when players' values are picked from the same distribution every round, liquid welfare is within a $2$ factor of the optimal one (up to an additive error, sublinear in the number of rounds) both in second and first-price auctions.
\citet{DBLP:journals/corr/GaitondeLLLS22} also prove that the weaker behavioral assumption that players have no-regret or a small competitive ratio is not enough to bound liquid welfare in second-price auctions: even when players have no-regret, the resulting liquid welfare can be arbitrarily bad compared to the optimal one.
In contrast, we prove such guarantees are possible in the case of first price.

In the offline setting, \citet{DBLP:journals/mansci/ConitzerKPSMSW22}, and more recently \citet{DBLP:journals/mansci/BalseiroKK23}, focus on shading-based equilibria in budgeted first-price auctions.
\citet{DBLP:journals/mansci/BalseiroKK23} prove that when every player's type is sampled from the same distribution and players shade their values and use them to bid according to standard symmetric first-price auction equilibria, then this produces a symmetric Bayesian Nash equilibrium of a single item auction while observing the budget limit in expectation.
They consider this soft budget limit (in expectation only) as when such an equilibrium solution is repeated many times, concentration will help essentially observe the true overall budget.
Their work naturally extends value shading, which is one of the several ways budgets are managed in practice, e.g., see \citep{DBLP:journals/ior/ConitzerKSM22,DBLP:journals/ior/BalseiroKMM21} for first-price and \citep{DBLP:journals/mansci/ConitzerKPSMSW22} for second-price.

Similar bounds for value-maximizing players with return-on-spend constraints are proven by \citet{DBLP:conf/wine/AggarwalBM19}, as well as by \citet{DBLP:conf/innovations/Babaioff0HIL21} for more general utility measures.
Both show that the liquid welfare of a pure Nash equilibrium of the static game is within a factor of $2$ of the optimal liquid welfare, when the underlying auction is truthful, e.g., a second-price auction.
\citet{DBLP:conf/www/DengMMZ21} improve this guarantee by using implicit or explicit information about the players' values.
In contrast, our efficiency results apply to player behavior in a repeated auction setting not only at equilibrium but also without converging to equilibrium.
This type of result was very recently found by \citet{DBLP:journals/corr/LucierPSZ23} for value-maximizing players under return-on-investment constraints using a specific algorithm: they come up with an adaptive pacing algorithm that when \textit{employed by every player} guarantees no-regret and the resulting liquid welfare is within a factor of $2$ of the optimal one which generalizes the result of \cite{DBLP:journals/corr/GaitondeLLLS22}.

\citet{DBLP:conf/stoc/SyrgkanisT13,DBLP:conf/soda/LemeST12} study social welfare inefficiencies for sequential first-price auctions when players have submodular valuations, no budget constraints, and are in equilibrium: for unit-demand valuations they prove a Price of Anarchy bound of $2$ while for general submodular valuations they prove that the Price of Anarchy can be unbounded.
We note that the latter result does not contradict our bound for submodular valuations: no-regret learning, while a useful behavioural model, does not lead to equilibrium behavior in the sequential game.
To the best of our knowledge, our result is the first positive welfare result in sequential auctions when players have general submodular valuations.
For simultaneous auctions, \cite{DBLP:conf/stoc/SyrgkanisT13} prove multiple Price of Anarchy bounds for social welfare.
Most relevant to our works are the ones for XOS valuations (a superset of submodular valuations): a $1.58$ Price of Anarchy bound in first-price and $2$ in second-price.
\cite{DBLP:journals/geb/FeldmanFGL20} prove a similar $2$ bound for first-price and $4$ for second-price when players have subadditive valuations.

Liquid welfare was introduced by \citet{DBLP:conf/icalp/DobzinskiL14} and was also used by \citet{DBLP:conf/stoc/SyrgkanisT13} (the latter refers to liquid welfare as effective welfare).
In terms of budgeted players, \citet{DBLP:conf/stoc/SyrgkanisT13} compare the resulting social welfare with the optimal liquid welfare, offering a rather unfair comparison, as the former can be arbitrarily bigger than the latter.
\citet{DBLP:conf/icalp/DobzinskiL14} focuses on designing mechanisms that maximize the liquid welfare.
\citet{DBLP:conf/aaai/FotakisLP19} convert known incentive compatible mechanisms that maximize social welfare for submodular players to incentive compatible mechanisms that maximize liquid welfare with similar guarantees.

Analyzing the outcome of regret-minimizing players in auctions has recently received attention from the research community.
\citet{DBLP:conf/www/KolumbusN22} study repeated auctions with two players who report their (potentially different) value to a regret-minimizing algorithm.
They notice that in second-price auctions the dynamics induced might not converge to the bidding of the equilibrium and therefore players might have incentive to misreport their values to the algorithms.
They also prove that this is not the case for first-price auctions.
Another similar line of work is that of \citet{DBLP:conf/aaai/0004GLMS21,DBLP:conf/www/DengHLZ22}, who study the convergence to equilibria of players who bid according to mean-based learning algorithms, a category of no-regret learning algorithms.
\citet{DBLP:conf/aaai/0004GLMS21} focuses on second-price, first-price, and multi-position VCG auctions when players have i.i.d. distributions and prove that the bids converge to the canonical Bayes-Nash equilibrium.
\citet{DBLP:conf/www/DengHLZ22} focuses on first-price auctions when players have fixed but different values and study conditions on the players' values that guarantee or not the convergence to a Nash equilibrium.

Another interesting series of related works is that of adversarial Bandits with Knapsacks.
Here, a budget-limited player tries to maximize her total reward by picking an action each round.
Each action has a reward and multiple costs, where each cost is subtracted from the budget of one of her resources; rewards and costs are picked by an adversary.
The framework was first introduced by \citet{DBLP:journals/jacm/ImmorlicaSSS22}, who prove that the player can achieve $O(d\log T)$ competitive ratio and sublinear regret, both in expectation and high probability, where $d$ is the number of resources.
\citet{DBLP:conf/colt/Kesselheim020} offer an improved $O(\log d \log T)$ competitive ratio.
\citet{DBLP:conf/icml/CastiglioniCK22} offer a $T/B$ competitive ratio guarantee (that matches the one by \cite{DBLP:journals/mansci/BalseiroG19} in second-price auctions) which improves the two previous results when the player's budget is linear in time, that is $T/B = O(1)$.
\citet{DBLP:conf/colt/FikiorisT23} prove that this competitive ratio is tight when $T/B = O(1)$.
Another earlier line of work focuses on stochastic BwK, where the rewards and costs are sampled from an unknown (potentially correlated) distribution every round.
In contrast to the adversarial setting, a competitive ratio of $1$ is achievable \citep{DBLP:journals/jacm/BadanidiyuruKS18}.
\citet{DBLP:conf/icml/CastiglioniCK22} provide an algorithm that simultaneously achieves competitive ratios of $1$ and $T/B$ in stochastic and adversarial BwK, respectively.
\citet{DBLP:conf/colt/FikiorisT23} generalize this result when the instance is between stochastic and adversarial.
An excellent discussion of both BwK and multi-armed bandits (BwK without constraints) can be found in \citet{DBLP:journals/ftml/Slivkins19}.

\section{Preliminaries and Model}

We assume there are $n$ players, $T$ rounds, and one item per round.
Every player\footnote{We denote $[n] = \{1,2,\ldots, n\}$ for $n \in \N$.} $i\in [n]$ has an additive valuation (we generalize this in \cref{sec:subm}): every round $t\in [T]$ she has a value $v_{it}$ for the item being auctioned that round and if she gets allocated the items of rounds $\mathcal T \sub [T]$ her total allocated value is
\begin{equation*}
    V_i = \sum_{t\in \mathcal T}v_{it}.
\end{equation*}
We assume that the players' values are bounded: $v_{it}\in [0,1]$ for every $i\in [n]$ and $t\in [T]$.

Unless stated otherwise, we focus on first-price auctions: in every round $t$, each player $i$ submits a bid $b_{it}$ and the player with the highest bid wins the item and pays her bid (ties are broken arbitrarily).
We denote with $p_t$ the price of the item, i.e., the highest bid, and with $d_{it}$ the highest competing bid faced by player $i$, i.e., $d_{it} = \max_{j\ne i}b_{jt}$.
If player $i$ gets allocated the items of rounds $\mathcal T \sub [T]$, then her total payment is
\begin{equation*}
    P_i = \sum_{t\in \mathcal T} p_t.
\end{equation*}

We assume that every player $i$ has a budget $B_i$ and \textit{budgeted quasi-linear utility}, i.e., her utility when her total value and payment are $V_i$ and $P_i$ is
\begin{equation*}
    U_i = 
    \begin{cases}
        V_i - P_i, &\textrm{ if } P_i\le B_i \\
        -\infty, &\textrm{ otherwise}
    \end{cases}
    .
\end{equation*}

We evaluate the auction system by measuring how well it does at maximizing the \textit{liquid welfare}: the contribution of player $i$ to the liquid welfare is
\begin{equation*}
    \lw_i = \min\left\{B_i, V_i\right\}
\end{equation*}
and the resulting liquid welfare is $\lw = \sum_{i\in [n]} \lw_i$.
We denote the optimal liquid welfare $\lw^* = \sum_i\lw_i^*$.
Additionally, for a subset of players $X\sub [n]$ we denote $\lw_X = \sum_{i\in X} \lw_i$ and similarly for $\lw_X^*$, $P_X$, etc.

\subsection{The behavioral assumption}

We assume all players learn to bid while participating in the sequential auctions.
When budgeted players are participating in repeated auctions where the values and prices may be adversarially picked, no-regret learning is not possible, but learning algorithms can achieve a bounded competitive ratio.
Specifically in this work, we compare a player's resulting utility with the following benchmark: her utility had she bid $\l$ times her value every round, up to her budget, for any $\l\in [0,1]$.
To formally define this benchmark, we first define $\hat U_i(\l)$ as the resulting utility of player $i$ if she had bid using shading multiplier $\l\in [0,1]$ every round, constrained by her budget.
Formally, $\hat U_i(\l)$ is the utility of player $i$ if her bid on every round $t$ was
\begin{equation*}
    \hat b_{it}(\l) = \min\left\{
            \l v_{it}, B_i - \sum_{\tau=1}^{t-1} \hat b_{i\tau}(\l) \One{\textrm{bid $\hat b_{i\tau}$ wins against $d_{i\tau}$}}
    \right\}
\end{equation*}
i.e., every round $t$ player $i$'s bid is the minimum of $\l v_{it}$ and her remaining budget.
Our results hold for any tie-breaking rule, which is why we do not explicitly state one above.
Using the above definition we define
\begin{equation*}
    \hat U_i(\l)
    =
    \sum_{t = 1}^T \left( v_{it} - \hat b_{it}(\l) \right) \One{\textrm{bid $\hat b_{it}$ wins against $d_{it}$}}
\end{equation*}
We emphasize that $d_{it}$, the highest competing bid faced by player $i$, is calculated using the bids that the other players used originally, not what their bids would have been if player $i$ had used $\hat b_{it}(\l)$.

Given the above definitions, the benchmark with which we compare a player $i$'s utility is the best-fixed shading multiplier, i.e., $\sup_{\l\in [0,1]}\hat U_i(\l)$.
We say that player $i$ has competitive ratio at most $\g \ge 1$ and regret at most $\reg$, if for her resulting utility $U_i$ it holds that
\begin{equation*}
    U_i \ge \frac{\sup_{\l\in [0,1]}\hat U_i(\l) - \reg}{\g}.
\end{equation*}
In the special case that $\g = 1$ and $\reg = o(T)$, we say that player $i$ has no-regret.

We note that the benchmark we use to compare player $i$'s resulting utility is much weaker than the one used in previous work, namely by \cite{DBLP:journals/mansci/BalseiroG19}.
More specifically, the benchmark they use is the \textit{best sequence of bids}, which is preferable in terms of individual guarantees.
In contrast, our liquid welfare guarantee is achieved by requiring the less restrictive behavior described above.
\section{Liquid Welfare Guarantees}\label{sec:guar}

In this section, we prove the guarantee for liquid welfare when all the players have a bounded competitive ratio.
We start with a lower bound for the benchmark we use to compare a player's utility, i.e., the utility she gets when she plays according to the optimal in retrospect shading multiplier.
We lower bound this using the player's contribution to the optimal liquid welfare.
Our bounded competitive ratio assumption and this bound will prove that players' utilities are also close to their contribution to the optimal liquid welfare.

\begin{lemma}\label{lem:guar:utility}
    Fix a player $i$, her values $v_{i1},\ldots,v_{iT}$, and the price of the items $p_1,\ldots,p_T$.
    Let $\calO_i\sub [T]$ be the items that player $i$ gets in the allocation that maximizes the total liquid welfare.
    Let $c(\l) = \frac{-1}{\ln(1-\l)}$.
    Then for any $\l\in(0,1)$, it holds that
    \begin{equation*}
        \sup_{\mu \in [0,1]} \hat U_i(\mu)
        \ge
        \min\left\{
            \frac{1-\l}{\l} \lw_i^*
            \,,\,
            c(\l)\l \lw_i^* - c(\l) \sum_{t\in\calO_i} p_t
        \right\}
         - 1.
    \end{equation*}
\end{lemma}

To prove the lemma we examine $\hat U_i(\l)$ (what happens when player $i$ uses multiplier $\l$) and distinguish two cases.
First, if using $\l$ makes player $i$ budget-constrained in any round, then she has spent almost all her budget (up to $\l$).
This lower bounds her utility since every time she wins an item the value she earns from it is at least $1/\l$ times the payment (regardless of being budget-constrained).
Second, if the player never becomes budget-constrained with multiplier $\l$, she is not constrained by any $\mu \le \l$.
This allows us to pick a random multiplier from a distribution and study the player's utility while ignoring the budget constraint.

\begin{myproof}
    Fix $\l$ and $\calO_i$ as described above.
    First, we examine the case where, if the player had used multiplier $\l$ to bid, then she runs out of budget, i.e., at some round $t$ she is budget constrained and therefore bids less than $\l v_{it}$.
    In this case, her total payment is at least $B_i - \l$ and every time she gets an item the value she gets from it is at least $1/\l$ times the price.
    This proves that in this case
    \begin{equation*}
        \hat U_i(\l)
        \ge
        \left( \frac{1}{\l} - 1 \right)(B_i - \l)
        =
        \left( \frac{1}{\l} - 1 \right)B_i - 1 + \l
    \end{equation*}
    which proves the lemma since $B_i \ge \lw_i^*$.
    
    Now we examine the case where player $i$ would not be budget constrained when using multiplier $\l$, which also implies she is not budget constrained for any multiplier $\mu\le\l$.
    If player $i$ uses multiplier $\mu \in [0,\l]$, then her utility is
    \begin{equation}\label{eq:guar:10}
        \hat U_i(\mu)
        \ge
        (1-\mu) \sum_{t\in[T]} v_{it} \One{\mu v_{it} > p_t}
        \ge
        (1-\mu) \sum_{t\in\calO_i} v_{it} \One{\mu v_{it} > p_t}
    \end{equation}
    where the first inequality is true because the highest bid that player $i$ faces in round $t$ is at most $p_t$.

    If the multiplier $\mu$ is picked from the distribution that has probability density function
    \begin{equation*}
        f_\l(\mu) =
        \begin{cases}
            \frac{c(\l)}{1-\mu}, & \text{ if }\mu\in [0,\l] \\
            0, & \textrm{ otherwise}
        \end{cases}
    \end{equation*}
    then taking the expectation of \eqref{eq:guar:10} we get
    \begin{alignat*}{4}
        & \Ex[\mu\sim f_\l(\mu)]{\hat U_i(\mu)}
        && \;\ge\; &&
        \int_{\mu=0}^\l (1-\mu) \sum_{t\in\calO_i} v_{it} \One{\mu v_{it} > p_t} \frac{c(\l)}{1-\mu} d\mu \\
        & && \;\ge\; &&
        c(\l) \sum_{t\in\calO_i} v_{it} (\l - p_t/v_{it}) \\
        & && \;=\; &&
        c(\l)\l \sum_{t\in\calO_i} v_{it} - c(\l) \sum_{t\in\calO_i}  p_t
        .
    \end{alignat*}

    The above proves what we want because $\sup_{\mu \in [0,1]} \hat U_i(\mu) \ge \Ex[\mu]{\hat U_i(\mu)}$ and $\sum_{t\in\calO_i}v_{it} \ge \lw_i^*$.
\end{myproof}

We now prove the guarantee for the total liquid welfare, given a bound for every player's competitive ratio.
We require the bound for the competitive ratio to be one with high probability (see \cref{thm:algo} for the guarantee on the competitive ratio we prove).

\begin{theorem}\label{thm:guar}
    Assume that every player $i$ has competitive ratio at most $\g\ge 1$ and regret $\reg$ with probability at least $1 - \d$.
    Then
    \begin{equation*}
        \lw
        \ge
        \frac{\lw^* - O(n) (\reg + 1)}{\g + 1/2 + O(1/\g)}
    \end{equation*}
    with probability at least $1-n\d$, for any $\d > 0$.
\end{theorem}

\begin{remark}
    As shown in the proof below, for $\g \le 1.73$ the Price of Anarchy of the above bound is $1 + \sqrt{\g + 1}$, which equals about $2.41$ for $\g = 1$.
\end{remark}

The theorem requires a competitive ratio with high probability and not in expectation.
This is because of the definition of liquid welfare: having high value with high probability implies high liquid welfare, but high expected value does not imply high liquid welfare.
We prove the theorem by partitioning the players into disjoint groups, depending on their value (if it is greater or not than their budget) and their utility (which of the two bounds of \cref{lem:guar:utility} holds).
After that, we pick the correct value of $\l$ that proves the bound.

\begin{myproof}
    Fix a player $i$ and recall the notation that $V_i$, $U_i$, and $P_i$ are $i$'s resulting value, utility, and payment, respectively.
    If player $i$ has competitive ratio at most $\g$ and regret at most $\reg$, it holds that
    \begin{alignat}{4}
        &V_i
        && \;=\; &&
        U_i + P_i
        \nonumber\\
        & && \;\ge\; &&
        \frac{1}{\g}\sup_{\mu \in [0,1]} \hat U_i(\mu) - \frac{\reg}{\g} + P_i
        && \qquad \textrm{(Competitive Ratio)}
        \nonumber\\
        & && \;\ge\; &&
        \frac{1}{\g}\min\left\{
            \frac{1-\l}{\l} \lw_i^*
            \,,\,
            c(\l)\l \lw_i^* - c(\l) \sum_{t\in\calO_i} p_t
        \right\}
        - \frac{\reg + 1}{\g} + P_i
        && \qquad \textrm{(\cref{lem:guar:utility})}
        .
        \label{eq:guar:9}
    \end{alignat}
    
    Now we partition the players into 3 sets:
    \begin{itemize}
        \item $X = \{i : V_i > B_i\}$.
        \item $Y = \{i : V_i \le B_i, \frac{1-\l}{\l} \lw_i^* \le c(\l)\l \lw_i^* - c(\l) \sum_{t\in\calO_i} p_t\}$.
        \item $Z = \{i : V_i \le B_i, \frac{1-\l}{\l} \lw_i^* > c(\l)\l \lw_i^* - c(\l) \sum_{t\in\calO_i} p_t\}$.
    \end{itemize}
    
    Given this partition and that with probability at least $1 - n\d$ inequality \eqref{eq:guar:9} holds for all players in $Y$ and $Z$, we get three inequalities\footnote{Recall the notation $\lw_X = \sum_{i\in X}\lw_i$, $P_X = \sum_{i\in X}P_i$, etc.}:
    \begin{equation}\label{eq:guar:11}
        \lw_X
        =
        B_X
        \ge
        \lw_X^*,
    \end{equation}
    \begin{equation}\label{eq:guar:12}
        \lw_Y
        \ge
        \frac{1-\l}{\g\l} \lw_Y^*
        + P_Y
        - |Y|\frac{\reg + 1}{\g},
    \end{equation}
    \begin{equation}\label{eq:guar:13}
        \lw_Z
        \ge
        \frac{c(\l)\l}{\g} \lw_Z^*
        + P_Z
        - \frac{c(\l)}{\g} P_{[n]}
        - |Z|\frac{\reg + 1}{\g}
    \end{equation}
    where in the last inequality we used that $\sum_{i\in Z}\sum_{t\in\calO_i} p_t \le \sum_{t\in [T]}p_t = P_{[n]}$, which follows from the fact that $\calO_1,\ldots,\calO_n$ are disjoint.
    We now solve each one of \eqref{eq:guar:11}, \eqref{eq:guar:12}, and \eqref{eq:guar:13} for $\lw_{(\cdot)}^*$, split $P_{[n]} = P_X + P_Y + P_Z$, and add them to get
    \begin{align}\label{eq:guar:51}
        & \lw^*
        -
        \max\left\{\frac{\l}{1-\l} , \frac{1}{c(\l)\l}\right\} n
        \left(\reg + 1\right)
        \nonumber\\
        &\;\le\;
        \lw_X + \frac{1}{\l} P_X
        +
        \frac{\g\l}{1-\l}\lw_Y + \left( \frac{1}{\l} - \frac{\g\l}{1-\l} \right)P_Y
        +
        \frac{\g}{c(\l)\l}\lw_Z + \left( \frac{1}{\l} - \frac{\g}{c(\l)\l} \right) P_Z
        .
    \end{align}

    We now use the fact that for all $i$, $0 \le P_i \le \lw_i + \reg/\g$, which follows from $P_i \le B_i$ (otherwise player $i$ would have unbounded negative utility) and $V_i - P_i \ge - \reg/\g$ (because of the competitive ratio guarantee and that for any $\mu$, $\hat U(\mu) \ge 0$).
    This proves that
    \begin{align}\label{eq:guar:52}
        & \frac{1}{\l} P_X
        +
        \left( \frac{1}{\l} - \frac{\g\l}{1-\l} \right) P_Y
        +
        \left( \frac{1}{\l} - \frac{\g}{c(\l)\l} \right) P_Z
        \nonumber\\
        &\;\le\;
        \frac{1}{\l} \left( \lw_X + |X| \frac{\reg}{\g}\right)
        +
        \left( \frac{1}{\l} - \frac{\g\l}{1-\l} \right)^+ \left( \lw_Y + |Y| \frac{\reg}{\g}\right)
        +
        \left( \frac{1}{\l} - \frac{\g}{c(\l)\l} \right)^+ \left(\lw_Z  + |Z| \frac{\reg}{\g}\right)
        \nonumber\\
        &\;\le\;
        \frac{1}{\l} \lw_X
        +
        \left( \frac{1}{\l} - \frac{\g\l}{1-\l} \right)^+ \lw_Y
        +
        \left( \frac{1}{\l} - \frac{\g}{c(\l)\l} \right)^+ \lw_Z
        +
        \frac{n}{\g \l}\reg
    \end{align}
    where in the last inequality we aggregated all the regret terms and used $\left( \frac{1}{\l} - \frac{\g\l}{1-\l} \right)^+ \le \frac{1}{\l}$ and $\left( \frac{1}{\l} - \frac{\g}{c(\l)\l} \right)^+ \le \frac{1}{\l}$.
    Plugging \eqref{eq:guar:52} into \eqref{eq:guar:51} we get
    \begin{align*}
        & \lw^*
        -
        \max\left\{\frac{\l}{1-\l} , \frac{1}{c(\l)\l}\right\} n
        \left(\reg + 1\right)
        -
        \frac{n}{\g\l} \reg
        \\
        &\;\le\;
        \left( 1 + \frac{1}{\l} \right)\lw_X
        +
        \left(\frac{\g\l}{1-\l} + \left( \frac{1}{\l} - \frac{\g\l}{1-\l} \right)^+ \right) \lw_Y
        +
        \left(\frac{\g}{c(\l)\l} + \left( \frac{1}{\l} - \frac{\g}{c(\l)\l} \right)^+ \right)\lw_Z
        \\
        &\;=\;
        \left( 1 + \frac{1}{\l} \right)\lw_X
        +
        \max\left\{\frac{\g\l}{1-\l} , \frac{1}{\l} \right\} \lw_Y
        +
        \max\left\{\frac{\g}{c(\l)\l} , \frac{1}{\l} \right\} \lw_Z
        \\
        &\;\le\;
        \max\left\{1 + \frac{1}{\l}, \frac{\g\l}{1-\l} , \frac{\g}{c(\l)\l} \right\} \lw
        .
    \end{align*}

    We now set an appropriate $\l$ in the above inequality.
    First, if $\g \le 1.73$, we set $\l = \l_1 = \frac{1}{\sqrt{1 + \g}}$ (which we get by solving $1 + \frac{1}{\l} = \frac{\g\l}{1-\l}$) and get
    \begin{equation*}
        1 + \frac{1}{\l_1}
        =
        \frac{\g\l_1}{1-\l_1}
        =
        1 + \sqrt{1 + \g}
        \ge
        \frac{\g}{c(\l_1)\l_1}
    \end{equation*}
    where one can prove that the inequality holds for $\g \le 1.73$ (see \cref{sec:app:calc}).

    For $\g \ge 1.73$, we solve the equation $1 + \frac{1}{\l} = \frac{\g}{c(\l)\l}$, set $\l_2 = \gamma  W\left(-\frac{e^{-2/\gamma }}{\gamma }\right)+1$, where $W(\cdot)$ is the \textit{Lambert W function}\footnote{The \textit{Lambert W function} $W(x)$ is the solution to the equation $y e^y = x$, \url{https://en.wikipedia.org/wiki/Lambert_W_function}.}, and get
    \begin{equation*}
        1 + \frac{1}{\l_2}
        =
        \frac{\g}{c(\l_2)\l_2}
        =
        1 + \frac{1}{\gamma  W\left(-\frac{e^{-2/\gamma }}{\gamma }\right)+1}
        \ge
        \frac{\g\l_2}{1-\l_2}
    \end{equation*}
    where one can prove that the inequality holds for $\g \ge 1.73$ (see \cref{sec:app:calc} along with a proof that $0 < \l_2 < 1$).
    Combining the two results we prove the theorem after showing that $1 + \frac{1}{\gamma  W\left(-\frac{e^{-2/\gamma }}{\gamma }\right)+1} = \g + 1/2 + O(1/\g)$ and the factor in front of the regret term is $O(n)$ (see \cref{sec:app:calc} for the proof of both).
\end{myproof}
\section{Impossibility Results for Liquid Welfare} \label{sec:imp}

In this section, we prove upper bounds for the guaranteed liquid welfare when the players have a competitive ratio of $\g$.
For completeness, we first include the bound for sequential budgeted second-price auctions of \cite{DBLP:journals/corr/GaitondeLLLS22}.

\begin{theorem}[{\citet[Proposition~D.1]{DBLP:journals/corr/GaitondeLLLS22}}]\label{thm:second_price}
    In sequential budgeted second-price auctions, for any number of rounds $T$ there exists an instance with two players where both have competitive ratio $1$ and $0$ regret and the resulting liquid welfare is arbitrarily smaller than the optimal one.
\end{theorem}

\begin{myproof}
    The first player has budget $B_1 = T\e$ for a small $\e>0$ and value $v_{1t} = 1$ every round.
    The second one has budget $B_2 = T$ and value $v_{2t} = 1$ every round.

    The optimal allocation is to give the item to the second player every round, achieving $\lw^* = T$.

    If player $1$ bids $b_{1t} = 1$ every round and player $2$ bids $b_{2t} = 0$, then the resulting liquid welfare is $\lw = T\e$ and every player has $0$ regret.
    Specifically, player $1$ gets every item for free, while player $2$ has no incentive to get any item at price $1$.
    This completes the proof by taking $\e\to 0$, since $\lw/\lw^* = \e$.
\end{myproof}

We now provide a liquid welfare upper bound for first-price auctions when every player has competitive ratio $\g = 1$.
Specifically, we prove that the liquid welfare can be half the optimal one, even when the players' values are constant across rounds and their no-regret guarantee is with respect to the best sequence of bids.
The last assumption shows that even if we strengthen the behavior assumption that we use in \cref{thm:guar} (players have high utility with respect to the best shading multiplier), we cannot guarantee that the liquid welfare is more than half the optimal one.

\begin{theorem}\label{thm:imp:2}
    In sequential budgeted first-price auctions, for every number of rounds $T$, there exists an instance where there are $n = 2$ players who have competitive ratio $1$ and constant regret against the best sequence of bids and the resulting liquid welfare is arbitrarily close to $\frac{1}{2}$ times the optimal one.
\end{theorem}

\begin{myproof}
    Fix $n = 2$ players and a small constant $\e$ such that $\e \ge 1/T$.
    The first player has value $v_{1t} = 1$ every round and a total budget of $B_1 = T\e$.
    The second player has value $v_{2t} = \e$ every round and a total budget of $B_2 = T\e$.

    An allocation is to give player $1$ the item for the first $\lceil T\e \rceil$ rounds and player $2$ the item for the rest of the rounds.
    This allocation results in liquid welfare equal to
    \begin{equation*}
        \min\left\{T\e, \lceil T\e \rceil \cdot 1\right\}
        +
        \min\left\{T\e, (T - \lceil T\e \rceil ) \e \right\}
        \ge
        T ( 2 \e  - \e^2 ) - \e
    \end{equation*}
    which proves that the optimal liquid welfare is $\lw^* \ge T(2\e - \e^2) - \e$.

    In contrast, if in every round player $1$ bids $\e$ and player $2$ bids $\e - 1/T$, then all the items are allocated to player $1$.
    Both players have competitive ratio $\g = 1$ and player $2$ has $0$ regret, while player $1$ has small regret because she could have bid lower.
    More specifically, her regret is the difference in utilities had she gotten every item at price $\e - 1/T$:
    \begin{equation*}
        T \big( 1 - (\e - 1/T) \big)
        -
        T (1 - \e)
        =
        1
    \end{equation*}
    
    Player $1$ gets all the items making the resulting liquid welfare $\lw = \min\left\{T\e, T \cdot 1\right\} + \min\left\{T\e, 0 \right\} = T\e$.
    This proves the theorem by noticing 
    \begin{equation*}
        \frac{\lw}{\lw^*}
        \le
        \frac{1}{2 - \e - \frac{1}{T}}
    \end{equation*}
    which proves the theorem by using $1/T \le \e$ and taking $\e \to 0$.
\end{myproof}

We now provide a second liquid welfare upper bound, which for large $\g$ makes \cref{thm:guar} (and also \cref{thm:subm} that we present later) tight.
More specifically, we show that if players have competitive ratio $\g$ then the resulting liquid welfare can be $\g$ times less than the optimal one.
Similar to \cref{thm:imp:2} we assume that the players have liquid welfare with respect to the stronger benchmark of the best sequence of bids.

\begin{theorem}\label{thm:imp:gamma}
    In sequential budgeted first-price auctions, for every number of rounds $T$ and $\g \ge 1$, there is an instance where every player has competitive ratio at most $\g$ and constant regret against the best sequence of bids, and the resulting liquid welfare is almost $\g$ times smaller than the optimal one.
\end{theorem}

Intuitively, the theorem's proof is based on the fact that, if there is only one player who gets only a $1/\g$ fraction of $T$ identical items, then her competitive ratio is $\g$ and the liquid welfare is $\g$ times less than the optimal.

\begin{myproof}
    There are two players, neither of which is budget-constrained.
    Player $1$ has value $v_{1t} = 1$ every round and player $2$ has value $v_{2t} = \e = 1/T$ every round.
    The optimal liquid welfare is $\lw^* = T$, by giving all the items to player $1$.

    The players' bids are the following: for the first $T/\g$ rounds, player $1$ bids $\e$, and player $2$ bids $0$.
    For the rest of the rounds, player $1$ bids $0$ and player $2$ bids $\e^2$.
    Player $1$ has a total utility of $U_1 = (1-\e)T/\g$ and player $2$ has $U_2 = (\e - \e^2) T(1-1/\g)$.
    This outcome yields liquid welfare $\lw = T/\g + \e T(1-1/\g)$ which is $\frac{1}{\g} + \e\frac{\g - 1}{\g} \le \frac{1}{\g} + \e$ fraction of $\lw^*$.

    We are left to prove that the above outcome has competitive ratio at most $\g$ and regret less than $1$ for every player.
    The best bidding sequence for player $1$ would have resulted in getting the first $T/\g$ items for free and the rest for a price of $\e^2$ (note that this results in more utility than using any constant shading multiplier).
    This allocation yields utility $T/\g + (1-\e^2)T(1-1/\g) = T\big(1 - \e^2(1-1/\g)\big)$.
    One can prove that this yields competitive ratio $\g$ and regret less than $1$ for player $1$.

    For player $2$ the best allocation would have been the second batch of items for free.
    The utility in that case is $\e T (1-1/\g)$.
    This yields a competitive ratio of $1$ and regret $T \e^2 (1-1/\g) = \e (1-1/\g) \le 1$.
\end{myproof}

\section{Algorithm for Bounded Competitive Ratio}\label{sec:algo}

In this section, we prove that a player with budget $B$ can achieve competitive ratio $T/B$ and sublinear regret with high probability against the best shading multiplier in retrospect, as needed by \cref{thm:guar}.
Our guarantee holds for any values of that player and behavior of the other players, even if they are adversarially picked.
We note that classic work in online learning with constraints usually focuses on the best in retrospect distribution of actions the player could have taken.
However, our results in \cref{sec:guar} do not require this stronger benchmark.

Our algorithm will be based on the adversarial Bandits with Knapsacks (BwK) setting, first studied by \cite{DBLP:journals/jacm/ImmorlicaSSS22}.
Unlike our setting, the action space in BwK is a discrete set.
As a first step towards our reduction, we prove that the difference in using two very close shading multipliers results in a small additive error.
This proves that uniformly discretizing the action space $[0,1]$ entails a small additive error.
Then we reduce the problem of using a finite set of shading multipliers in sequential budgeted first-price auctions to the framework of adversarial BwK with small additive error.

We first prove that using shading multiplier $\l+\e$ instead of $\l$ yields a small additive error, proportional to $\e$.
We focus on a single arbitrary player and so we drop the $i$ subscript throughout this section.

\begin{lemma}\label{lem:algo:discretization}
    Fix any highest bids the other players have submitted, the player's budget $B \le T$, her values $v_1,\ldots, v_T$, and let $\hat U(\l)$ be the utility of the player had she used multiplier $\l\in [0,1]$.
    Then for any $\l\in[0,1]$ and $\e\in [0,1-\l]$ it holds that
    \begin{equation*}
        \hat U(\l+\e) \ge \hat U(\l) - \frac{T^2 \e}{B} - 2.
    \end{equation*}
\end{lemma}

The lemma is proven by examining two cases.
First, if using multiplier $(\l+\e)$ does not make the player run out of budget, then using that multiplier instead of $\l$ can only result in a slightly larger payment, at most $T\e$.
If by using multiplier $(\l+\e)$ the player runs out of budget, then her payment is almost her budget, which yields a high utility since the value of every item she gets is at least $\frac{1}{\l+\e}$ times the price she paid for it.
The second case is a bit more complicated: since the new multiplier might win arbitrarily more items, it might also result in running out of budget much faster than using multiplier $\l$.
This could result in the two multipliers winning arbitrarily different sets of items.

\begin{myproof}
    We first study the outcome when the player uses multiplier $\l$.
    Let $\calT\sub[T]$ be the rounds the player wins when biding with multiplier $\l$ and is not budget constrained, i.e., she bids $\l$ times her value and wins the item.
    If at any round the player wins an item while being budget-constrained, then she bids and pays her entire remaining budget.
    This means that, other than the items in $\calT$, the player wins at most one more item, whose value is at most $1$, proving that
    \begin{equation}\label{eq:algo:1}
        \hat U(\l)
        \le
        \sum_{t\in \calT}(1-\l)v_t + 1.
    \end{equation}
    
    When the player bids with multiplier $\l+\e$ then she is guaranteed to win every item of rounds $\calT$, unless she runs out of budget.
    This means that she either gets utility at least $(1-\l-\e)\sum_{t\in\calT} v_t$ or pays at least $B-(\l+\e)$.
    This implies that
    \begin{equation}\label{eq:algo:2}
        \hat U(\l+\e)
        \ge
        \begin{cases}
            (1-\l-\e)\sum_{t\in\calT} v_t,
            &\textrm{ if player gets $\calT$ without being budget constrained}
            \\
            \left(\frac{1}{\l+\e} - 1\right)\big(B-(\l+\e)\big),
            & \textrm{ if } (\l+\e)\sum_{t\in\calT} v_t \le B, \textrm{ but does not get $\calT$}
            \\
            \left(\frac{1}{\l+\e} - 1\right)\big(B-(\l+\e)\big),
            & \textrm{ if } (\l+\e)\sum_{t\in\calT} v_t > B
        \end{cases}
    \end{equation}
    where the second and third cases come from the fact that every time the player wins an item, her value for it is at least $\frac{1}{\l+\e}$ times the price she pays for it.

    For the second case of \eqref{eq:algo:2}, because $(\l+\e)\sum_{t\in\calT} v_t \le B$, we have
    \begin{equation*}
        \hat U(\l+\e)
        \ge
        \left(\frac{1}{\l+\e} - 1\right)\big(B-(\l+\e)\big)
        \ge
        \left(1 - \l -\e\right)\sum_{t\in\calT}v_t - 1 + \l + \e
        \ge
        \hat U(\l) -\e T - 1
    \end{equation*}
    where the last inequality holds because of \eqref{eq:algo:1}, $\sum_t v_t \le T$, and $\l+\e \ge 0$.
    Because $B \le T$, the inequality above satisfies the lemma and is also similarly proved for the first case of \eqref{eq:algo:2}.

    For the third case of \eqref{eq:algo:2}, we have that
    \begin{alignat*}{4}
        & \hat U(\l+\e)
        && \;\ge\; &&
        \left(\frac{1}{\l+\e} - 1\right) B - 1 + \l + \e
        \\
        & && \;\ge\; &&
        \frac{1-\l-\e}{\l+\e} \l \sum_{t\in\calT}v_t - 1
        && \qquad \left(\l \sum_{t\in\calT}v_t \le B \textrm{ and }\l+\e\ge 0\right)
        \\
        & && \;\ge\; &&
        \hat U(\l)
        + 
        \left(
            \frac{1-\l-\e}{\l+\e} \l -
            (1 - \l)
        \right)\sum_{t\in\calT}v_t - 2
        && \qquad \big(\textrm{using \eqref{eq:algo:1}}\big)
        \\
        & && \;=\; &&
        \hat U(\l)
        -
        \frac{\e}{\l + \e}\sum_{t\in\calT}v_t - 2
        \\
        & && \;\ge\; &&
        \hat U(\l)
        -
        \frac{T^2 \e}{B}  - 2
        && \qquad \left((\l+\e)\sum_{t\in\calT} v_t > B, \sum_{t\in\calT} v_t \le T\right)
    \end{alignat*}

    The above inequality satisfies the lemma and completes the proof.
\end{myproof}

Now we prove that given any discretization of the bid space $[0, 1]$, we can use any algorithm from BwK to achieve a low competitive ratio and no-regret in first-price auctions.
We first briefly present the BwK framework.

\begin{definition}[BwK framework]
    There are $K+1$ actions, $T$ rounds, and a resource with a total budget of $B$.
    The adversary picks rewards $(r_{t,k})_{t\in [T], k\in [K]}$ and costs $(c_{t,k})_{t\in [T], k\in [K]}$ for every round-action pair.
    The $0$-th arm is assumed to have $c_{t,0} = r_{t,0} = 0$.
    On round $t$, without observing the rewards and costs of that round, the player picks a (potentially randomized) action $k_t = 0, \ldots, K$.
    The game ends after round $T$ or on round $T'$ when the player depletes the resource, i.e., when $\sum_{t=1}^{T'}c_{t,k_t} > B$.
    We denote the total reward of the player with $\rew = \sum_{t=1}^{\min\{T, T'-1\}}r_{t,k_t}$ and with $\optfa$ the reward of best-fixed action in retrospect.
    We say that the player has competitive ratio $\g$ and regret at most $\reg$ if
    \begin{align*}
        \rew
        \ge
        \frac{\optfa - \reg}{\g}.
    \end{align*}

    If the adversary samples the rewards and costs of every round from a distribution independent of other rounds, the setting is \textit{stochastic}.
    If the adversary picks them arbitrarily before the first round (i.e., without seeing any of the player's actions), then the adversary is called \textit{oblivious}.
    Finally, if they are picked with knowledge of previous rounds the adversary is called \textit{adaptive}.
\end{definition}

\begin{lemma}\label{lem:algo:reduction}
    Let $\l_1,\ldots,\l_K \in [0,1]$ be any multipliers and $\l_0 = 0$.
    Using any algorithm for BwK that with probability $1-\d$ has competitive ratio $\g$ and regret $\reg_T(\d, K)$ against an adaptive adversary, we get an algorithm for repeated first-price auctions that uses only multipliers $\l_0,\ldots,\l_K$ and achieves utility $U$ for which
    \begin{equation*}
        \Pr{
        U
        \ge
        \frac{\max_{k=0,\ldots,K}\hat U(\l_k) - \reg_T(\d, K) - 2}{\g}
        } \ge 1 - \d
    \end{equation*}
    even if an adaptive adversary picks the values and prices of the items.
\end{lemma}

\begin{myproof}
    To reduce the problem of learning in sequential first-price auctions to BwK, we set\footnote{Here we make the pessimistic assumption that the player needs to bid strictly above the highest competing bid to win an auction but our results hold for any tie braking rule, even if it changes across rounds.} $r_{t,k} = (1-\l_k) v_t\One{\l_k v_t > d_t}$ and $c_{t,k} = \l_k v_t\One{\l_k v_t > d_t}$.
    We then run the algorithm from the BwK setting on these rewards and costs.
    
    First, we note a small mismatch between the BwK and the repeated auction settings.
    A sequence of actions has the same rewards, costs, and remaining budget in both settings, up to the round where in the BwK setting the player runs out of budget.
    In BwK if the player picks an action that incurs a cost higher than the remaining budget the game stops.
    In contrast, in the repeated auction setting, if the player's bid is higher than her remaining budget, then her bid is adjusted, which may or may not end the game by depleting her budget.
    However, this causes a small mismatch:
    \begin{itemize}
        \item For the difference between the reward of the algorithm in BwK and the utility of the same actions in the repeated auction setting (where the actions after the BwK algorithm runs out of budget are picked arbitrarily), it holds that $U \ge \rew$.
        
        \item For the two benchmarks, $\optfa$ and $\max_k \hat U(\l_k)$, it holds $\max_k \hat U(\l_k) \le \optfa + 2$.
        The rewards when playing the $k$-th arm and using multiplier $\l_k$ are the same up to before the round when the game stops in the BwK setting.
        On that round and onwards, in the auction setting the player has less than $\l_k$ budget remaining.
        As long as this remaining budget is used to win items without lowering the player's bid (i.e., the player bids $\l_k v_t$), the player can gain at most $1-\l_k$ additional utility, since the utility she gains is $\frac{1}{\l_k}-1$ times the price she pays and she has at most $\l_k$ remaining budget.
        Additionally, she might earn one more item on the round her budget is depleted.
        This means that after having less than $\l_k$ remaining budget, she may earn up to $1-\l_k + 1 \le 2$ utility.
    \end{itemize}
    
    Using the fact that the two above facts are true almost surely and that
    \begin{align*}
        \Pr{
            \rew
            \ge
            \frac{\optfa - \reg_T(\d, K)}{\g}
        } \ge 1 - \d
    \end{align*}
    we get the lemma.
\end{myproof}

Finally, we prove the main result of this section by combining the two previous lemmas and the algorithm of \citet{DBLP:conf/colt/FikiorisT23,DBLP:conf/icml/CastiglioniCK22}.
We use their result that the aforementioned algorithm achieves $T/B$ competitive ratio with high probability, even against an adaptive adversary.

\begin{theorem}\label{thm:algo}
    Fix a player with budget $B$.
    When after bidding the player gets to know the highest competing bid, there is an algorithm for sequential first-price auctions that for any $\d > 0$ achieves utility $U$ for which
    \begin{equation*}
        \Pr{
        U
        \ge
        \frac{\sup_{\l\in [0,1]}\hat U(\l) - O\left( \frac{T^{3/2}}{B} \sqrt{\log(T/\d)} \right)}{T/B}
        } \ge 1 - \d
    \end{equation*}
    even if an adaptive adversary picks the values and prices of the items.
\end{theorem}

Our result is only meaningful when $B = T^{1/2 + \Omega(1)}$, because otherwise, the regret is not sublinear.
This is reminiscent of the requirement of $B = \Omega(\sqrt T)$ that BwK algorithms have (there is also a $T^2/B$ lower bound on the competitive ratio when $B < \sqrt T$ see \cite{DBLP:journals/jacm/ImmorlicaSSS22}).
Additionally, our competitive ratio $T/B$ is constant when $B = \Omega(T)$, in which case it is tight; \cite{DBLP:journals/mansci/BalseiroG19} prove this for second-price auctions and their example can be extended to first-price as well.

\begin{myproof}
    Fix $K = T^2$ and let $\l_k = k/K$ for $k = 0,1,\ldots, K$.
    Using the algorithm of \cite{DBLP:conf/colt/FikiorisT23} that has competitive ratio $\g = T/B$ and regret at most $\reg_T(\d, K) = O\left( \frac{T}{B}\sqrt{T\log(T K/\d)} \right)$ with probability $1-\d$ and \cref{lem:algo:reduction} to get that with probability at least $1 - \d$,
    \begin{equation*}
        U
        \ge
        \frac{\max_{k=0,\ldots,K}\hat U(\l_k) - O\left( \frac{T}{B} \sqrt{T\log(T K/\d)} \right)}{T / B}
        =
        \frac{\max_{k=0,\ldots,K}\hat U(\l_k) - O\left( \frac{T^{3/2}}{B}\sqrt{\log(T/\d)} \right)}{T / B}
    \end{equation*}
    
    Fix any multiplier $\l \in [0,1]$ and let $\l_k$ be such that $\l \le \l_k \le \l - 1/K$.
    Using \cref{lem:algo:discretization} we get that $\hat U(\l_k) \ge \hat U(\l) - O(\frac{T^2}{K B}) - 2 = \hat U(\l) - O(1)$.
    This proves that $\max_k\hat U(\l_k) \ge \sup_\l\hat U(\l) - O(1)$ and completes the proof of the theorem.
\end{myproof}

\begin{remark}
    We note that even though \cref{thm:algo} proves a high probability $T/B$ competitive ratio and sublinear regret for sequential budgeted first-price auctions when $B = T^{1/2 + \Omega(1)}$, our proof can be easily adapted to use the guarantee of any BwK algorithm, either with high probability or in expectation, in the stochastic or adversarial case.
    For example, for certain values of $B$, we could get the $O(\log T)$ competitive ratio guarantee that \cite{DBLP:journals/jacm/ImmorlicaSSS22} achieve against an oblivious adversary.
    In addition, when the player has bandit feedback (i.e., only observes the value and payment of her action) then we can get the same competitive ratio with slightly higher regret.
\end{remark}

\section{Liquid Welfare Guarantees for Submodular Valuations}\label{sec:subm}

We now move to the final section of our results, where we generalize the results of \cref{sec:guar} for the case where players have submodular valuations.
If player $i$ receives a bundle of items $\calT\sub [T]$ then her value is $v_i(\calT)$, where $v_i$ is a submodular, non-decreasing, and non-negative function, i.e., for any subsets of rounds $\calS\sub\calT\sub [T]$ and $t\notin\calT$ it holds that $v_i(\calS\cup\{t\}) - v_i(\calS) \ge v_i(\calT\cup\{t\}) - v_i(\calT) \ge 0$ and $v_i(\calS) \ge 0$.
We use the standard notation for the marginal value of item $t$ for bundle $\calT$: $v_i\marg{t}{\calT} = v_i(\calT\cup\{t\}) - v_i(\calT)$.
Similar to the additive valuations setting, we normalize the players' valuations so that $v_i\marg{t}{\calT} \le 1$ for all items $t$ and bundles $\calT$.
The definitions of the players' utilities and liquid welfare remain the same.

Before bounding the liquid welfare in this setting, we first define how bidding according to a multiplier $\l\in [0,1]$ works in this case.
If player $i$ in round $t$ uses multiplier $\l$ and has already gained items $\calT\sub [t-1]$ then she bids $\l$ her current marginal value for item $t$, $\l v_i\marg{t}{\calT}$, as long as she is not budget constrained (if she is budget constrained she bids her remaining budget).
Because the marginal value of the item in every round (and therefore the bid) of every player depends on her past allocation, this setting is more complicated than the one we studied in previous sections.
Most notably, it is not clear if there exists an algorithm with bounded competitive ratio and regret, as we showed in \cref{sec:algo} for the additive case.
The reason is that a reduction to BwK is much harder since the reward and consumption of the resource in a single round depend on the results of the previous ones.
We leave the last question as future work.

In this section, we prove the following theorem, which proves a slightly worse bound than the one of \cref{thm:guar}.

\begin{theorem}\label{thm:subm}
    Assume that every player $i$ has competitive ratio at most $\g\ge 1$ and regret $\reg$ with probability at least $1 - \d$.
    If the players have submodular valuations then
    \begin{equation*}
        \lw
        \ge
        \frac{\lw^* - O(n) (\reg + 1)}{\frac{2 + \g + \sqrt{\g^2 + 4}}{2}}
    \end{equation*}
    with probability at least $1 - n\d$, for any $\d\in(0,1/n)$.
\end{theorem}

\begin{remark}
    We note that the factor in the denominator in \cref{thm:subm} for $\g = 1$ is about $2.62$ and for all $\g$ is a bit bigger than the one in \cref{thm:guar}.
    Both asymptotically are $\g + O(1)$.
    We give a plot of both for small $\g$ in \cref{fig:subm}.
    
    \begin{figure}[t]
        \centering
        \includegraphics[height=.22\textheight]{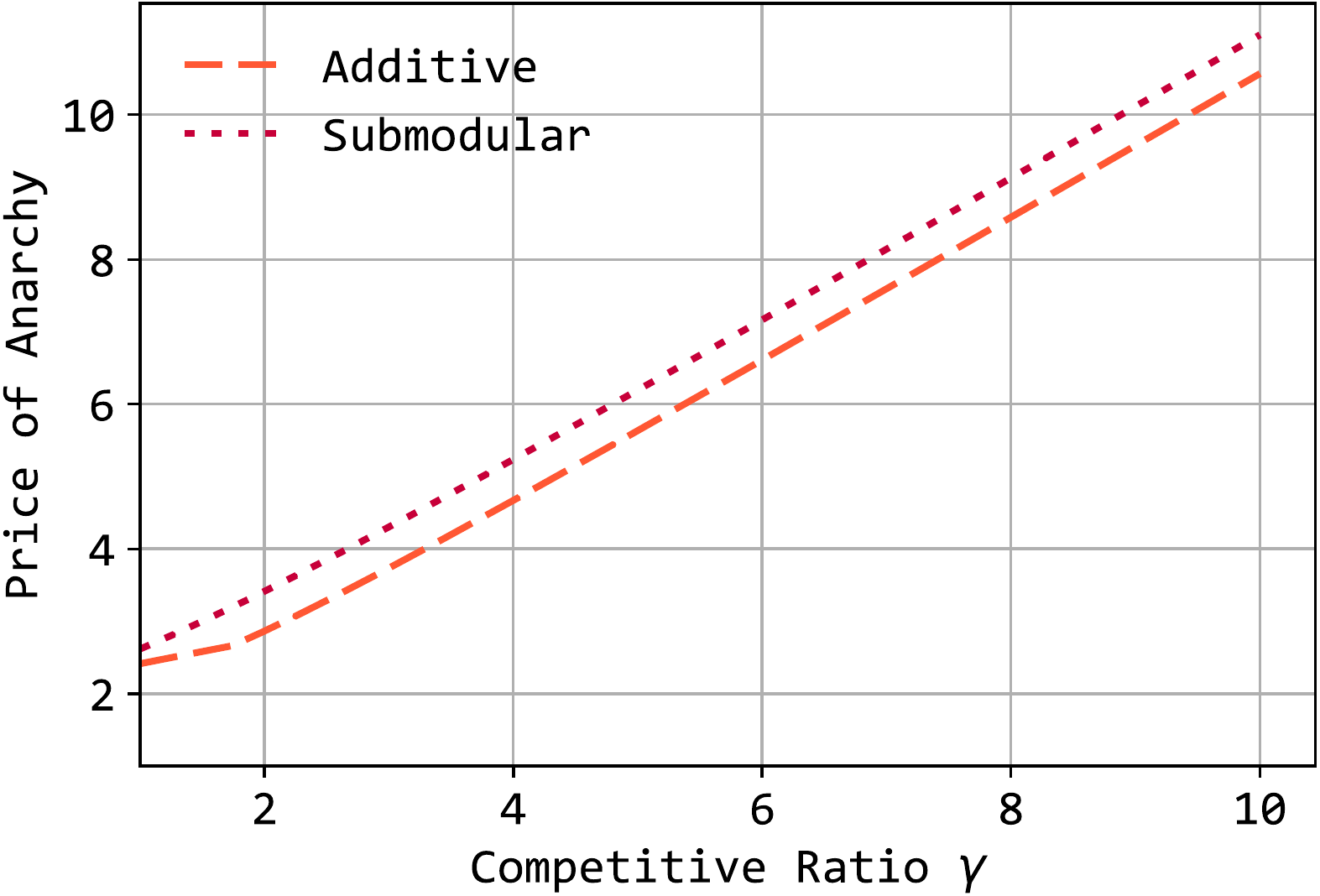}
        \caption{Price of Anarchy plots for Theorems \ref{thm:guar} (additive) and \ref{thm:subm} (submodular) for $\g\in [1, 10]$.}
        \label{fig:subm}
    \end{figure}
\end{remark}

We start with a simple lemma that will help us lower bound the value of the bundle $\calT_i$ gained by player $i$ when she uses a fixed shading multiplier $\l$.
More specifically, we know that if the player is not budget constrained in round $t$, then $t \notin \calT_i$ implies $\l v_i\marg{t}{\calT_i\cap [t-1]} \le p_t$.
In the following lemma, we lower bound $v_i(\calT_i)$ for any $\calT_i$ that satisfies the previous implication.

\begin{lemma}\label{lem:subm}
    Fix a player $i$, her submodular valuation $v_i:2^{[T]} \to \R_{\ge 0}$, and the resulting prices of the items $p_1,\ldots,p_T$.
    Let $\l\in(0,1]$ and $\calT_i\sub [T]$ such that
    \begin{equation*}
        t\notin \calT_i
        \implies
        v_i\marg{t}{\calT_i\cap [t-1]} \le \frac{1}{\l} p_t.
    \end{equation*}

    Then, for any set $\calO_i\sub [T]$ it holds that
    \begin{equation*}
        v_i(\calT_i)
        \ge
        v_i(\calO_i) - \frac{1}{\l}\sum_{t\in \calO_i} p_t.
    \end{equation*}
\end{lemma}

\begin{myproof}
    We have that
    \begin{alignat*}{4}
        & v_i(\calT_i) + \frac{1}{\l}\sum_{t\in \calO_i} p_t
        && \;\ge\; &&
        v_i(\calT_i) + \frac{1}{\l}\sum_{t\in \calO_i\setminus \calT_i} p_t 
        &&\qquad (p_t \ge 0)
        \\
        & && \;\ge\; &&
        v_i(\calT_i) + \sum_{t\in \calO_i\setminus \calT_i} v_i\marg{t}{\calT_i\cap [t-1]}
        &&\qquad (t\notin \calT_i)
        \\
        & && \;\ge\; &&
        v_i(\calT_i) + \sum_{t\in \calO_i\setminus \calT_i} v_i\marg{t}{\calT_i\cup(\calO_i\cap [t-1])}
        &&\qquad (\textrm{submodularity})
        \\
        & && \;=\; &&
        v_i(\calO_i \cup \calT_i) \ge v_i(\calO_i)
    \end{alignat*}
    where in the last equality, every term in the sum iteratively adds the marginal of an item $t$ that is in $\calO_i$ and not in $\calT_i$.
\end{myproof}

Next, we prove a lemma analogous to \cref{lem:guar:utility} and lower bound a player's utility for any fixed shading multiplier.
The following lemma provides a worse guarantee than \cref{lem:guar:utility} because it does not use randomization on the multiplier picked.

\begin{lemma}\label{lem:subm:utility}
    Fix a player $i$ and the resulting prices of the items $p_1, \ldots, p_T$.
    Then, for any $\l\in(0,1)$ it holds that
    \begin{equation*}
        \hat U_i(\l)
        \ge
        (1-\l)\left( \lw_i^* - \frac{1}{\l}\sum_{t\in \calO_i}p_t \right)
        -1
    \end{equation*}
    where $\calO_i \sub [T]$ is the bundle player $i$ gets in the allocation that maximizes the total liquid welfare.
\end{lemma}

\begin{myproof}
    Assume that by using multiplier $\l$, player $i$ would have gotten bundle $\calT_i$ if she was never budget constrained; in that case, she would have had utility $(1-\l)v_i(\calT_i)$ and it would have held $\l v_i(\calT_i) \le B_i$.
    If she was budget constrained (in which case $\l v_i(\calT_i) > B_i$) she would have spent at least $B_i - \l$, which yields a utility of at least $(1/\l - 1)(B_i - \l)$, since the marginal value she gets from any item she wins is at least $1/\l$ what she pays for it.
    This proves
    \begin{alignat*}{4}
        & \hat U_i(\l)
        && \;\ge\; &&
        \begin{cases}
            (1-\l) v_i(\calT_i), &\textrm{ if } \l v_i(\calT_i) \le B_i \\
            \left(\frac{1}{\l} - 1\right)(B_i - \l), &\textrm{ if } \l v_i(\calT_i) > B_i \\
        \end{cases}\\
        & && \;\ge\; &&
        (1-\l)\min\left\{ v_i(\calT_i) \;, \frac{1}{\l}B_i - 1 \right\}.
    \end{alignat*}

    The items not in bundle $\calT_i$ are the ones that multiplier $\l$ would not win, irrespective of whether the player becomes constrained by her budget.
    This entails that $\calT_i$ satisfies the requirements of \cref{lem:subm}, making the above
    \begin{equation*}
        \hat U_i(\l)
        \ge
        (1-\l)\min\left\{ v_i(\calO_i) - \frac{1}{\l}\sum_{t\in \calO_i}p_t \;, \frac{1}{\l}B_i - 1 \right\}.
    \end{equation*}
    
    The above proves what we want since $v_i(\calO_i) \ge \lw_i^*$, $p_t \ge 0$, and $\frac{1}{\l}B_i \ge \frac{1}{\l}\lw_i^* \ge \lw_i^*$.
\end{myproof}

We now prove \cref{thm:subm}, by picking the right multiplier $\l$.
The proof is similar to the one in \cref{thm:guar}.

\begin{myproof}[Proof of \cref{thm:subm}]
    Fix a player $i$ and recall the notation that $V_i$, $U_i$, and $P_i$ are $i$'s resulting value, utility, and payment, respectively.
    If player $i$ has competitive ratio at most $\g$ and regret at most $\reg$, it holds that
    \begin{alignat*}{4}
        &V_i
        && \;=\; &&
        U_i + P_i
        \\
        & && \;\ge\; &&
        \frac{1}{\g}\sup_{\mu \in [0,1]} \hat U_i(\mu) - \frac{\reg}{\g} + P_i
        && \qquad \textrm{(Competitive Ratio)}
        \\
        & && \;\ge\; &&
        \frac{1-\l}{\g}\left(
            \lw_i^* - \frac{1}{\l}\sum_{t\in \calO_i}p_t
        \right)
        - \frac{\reg + 1}{\g} + P_i
        && \qquad \textrm{(\cref{lem:subm:utility})}
    \end{alignat*}

    Now we partition the players into 2 sets:
    \begin{enumerate}
        \item $X = \{i : V_i > B_i\}$.
        \item $Y = \{i : V_i \le B_i\}$.
    \end{enumerate}
    
    Given this partition and that with probability at least $1 - n\d$ the above inequality holds for all players in $Y$, we get two inequalities\footnote{Recall the notation $\lw_X = \sum_{i\in X}\lw_i$, $P_X = \sum_{i\in X}P_i$, etc.}:
    \begin{equation}\label{eq:subm:11}
        \lw_X
        =
        B_X
        \ge
        \lw_X^*,
    \end{equation}
    \begin{equation}\label{eq:subm:12}
        \lw_Y
        \ge
        \frac{1-\l}{\g} \lw_Y^*
        + P_Y
        - \frac{1-\l}{\g \l} P_{[n]}
        - |Y|\frac{\reg + 1}{\g}
    \end{equation}
    where in the last inequality we used the fact that $\calO_1,\ldots,\calO_n$ are disjoint.

    Now solving \eqref{eq:subm:12} for $\lw_Y^*$ and adding it with \eqref{eq:subm:11} we get
    \begin{align*}
        & \lw^*
        -
        \frac{1}{1-\l} n
        \left(\reg + 1\right)
        \\
        &\;\le\;
        \lw_X + \frac{1}{\l} P_X
        +
        \frac{\g}{1-\l}\lw_Y + \left( \frac{1}{\l} - \frac{\g}{1-\l} \right) P_Y
    \end{align*}

    We now use the fact that $0 \le P_i \le \lw_i + \reg/\g$, which follows from $P_i \le B_i$  and $V_i - P_i \ge - \reg/\g$ (both following because player $i$ has bounded competitive ratio).
    This makes the above inequality
    \begin{align*}
        & \lw^*
        -
        \frac{1}{1-\l} n
        \left(\reg + 1\right)
        -
        \frac{n}{\g\l} \reg
        \\
        &\;\le\;
        \left( 1 + \frac{1}{\l} \right)\lw_X
        +
        \left(\frac{\g}{1-\l} + \left( \frac{1}{\l} - \frac{\g}{1-\l} \right)^+ \right)\lw_Y
        \\
        &\;=\;
        \left( 1 + \frac{1}{\l} \right)\lw_X
        +
        \max\left\{\frac{\g}{1-\l} , \frac{1}{\l} \right\} \lw_Y
        \\
        &\;\le\;
        \max\left\{1 + \frac{1}{\l}, \frac{\g}{1-\l} \right\} \lw
        .
    \end{align*}

    To get the desired bound, we set $\l$ appropriately in the above inequality.
    We set $\l = \frac{2}{\g + \sqrt{\g^2 + 4}}$, which we get by equalizing the two terms in the maximum: $1 + \frac{1}{\l} = \frac{\g}{1-\l}$.
    This entails $1 + \frac{1}{\l} = \frac{\g}{1-\l} = \frac{2 + \g + \sqrt{\g^2 + 4}}{2}$ and $\frac{1}{1 - \l} + \frac{1}{\g \l} = \frac{1 + \g + \sqrt{\g^2 + 4}}{\g}$, which makes the above inequality
    \begin{equation*}
        \lw^*
        -
        \frac{1 + \g + \sqrt{\g^2 + 4}}{\g} n \left(\reg + 1\right)
        \le
        \frac{2 + \g + \sqrt{\g^2 + 4}}{2} \lw
    \end{equation*}
    and proves the theorem since $\frac{1 + \g + \sqrt{\g^2 + 4}}{\g} \le 2+\sqrt 5$ for all $\g \ge 1$.
\end{myproof}

\section{Conclusion}

We perform an extensive study of liquid welfare guarantees in sequential budgeted first-price auctions.
We focus on first-price auctions since previous work shows that in second-price, even under strong conditions, liquid welfare guarantees are not implied by any no-regret guarantees.
In contrast, we show that for first-price auctions, any competitive ratio guarantee translates to liquid welfare guarantees, even when item values are adversarial and valuations are submodular.

While we prove positive liquid welfare guarantees under mild learning assumptions, we leave open the problem of whether stronger learning guarantees imply better liquid welfare.
For example, can we get better liquid welfare guarantees if the players have high competitive ratio with respect to the best Lipschitz continuous function that maps values to bids?
Or if the values are sampled from a distribution instead of being picked adversarially?
These questions are especially interesting in the regime where $\g$ is equal (or close) to $1$, where our lower and upper bounds are furthest apart, $2$ and $2.41$, respectively.

Another open line of work concerns the secondary focus of our work, learning in sequential budgeted auctions: is it possible for a player to guarantee low competitive ratio when her value is submodular?

\printbibliography{}

@article{alcobendas2021adjustment,
  title   = {Adjustment of Bidding Strategies After a Switch to First-Price Rules},
  author  = {Alcobendas, Miguel and Zeithammer, Robert},
  journal = {Available at SSRN 4036006},
  year    = {2021}
}

@article{choi2020online,
  title     = {Online display advertising markets: A literature review and future directions},
  author    = {Choi, Hana and Mela, Carl F and Balseiro, Santiago R and Leary, Adam},
  journal   = {Information Systems Research},
  volume    = {31},
  number    = {2},
  pages     = {556--575},
  year      = {2020},
  publisher = {INFORMS}
}

@inproceedings{DBLP:conf/aaai/FotakisLP19,
  author    = {Dimitris Fotakis and
               Kyriakos Lotidis and
               Chara Podimata},
  title     = {A Bridge between Liquid and Social Welfare in Combinatorial Auctions
               with Submodular Bidders},
  booktitle = {The Thirty-Third {AAAI} Conference on Artificial Intelligence, {AAAI}
               2019, The Thirty-First Innovative Applications of Artificial Intelligence
               Conference, {IAAI} 2019, The Ninth {AAAI} Symposium on Educational
               Advances in Artificial Intelligence, {EAAI} 2019, Honolulu, Hawaii,
               USA, January 27 - February 1, 2019},
  pages     = {1949--1956},
  publisher = {{AAAI} Press},
  year      = {2019},
  url       = {https://doi.org/10.1609/aaai.v33i01.33011949},
  doi       = {10.1609/aaai.v33i01.33011949},
  bibsource = {dblp computer science bibliography, https://dblp.org}
}

@inproceedings{DBLP:conf/colt/Kesselheim020,
  author    = {Thomas Kesselheim and
               Sahil Singla},
  editor    = {Jacob D. Abernethy and
               Shivani Agarwal},
  title     = {Online Learning with Vector Costs and Bandits with Knapsacks},
  booktitle = {Conference on Learning Theory, {COLT} 2020, 9-12 July 2020, Virtual
               Event [Graz, Austria]},
  series    = {Proceedings of Machine Learning Research},
  volume    = {125},
  pages     = {2286--2305},
  publisher = {{PMLR}},
  year      = {2020},
  url       = {http://proceedings.mlr.press/v125/kesselheim20a.html},
  bibsource = {dblp computer science bibliography, https://dblp.org}
}

@article{DBLP:journals/mansci/ConitzerKPSMSW22,
  author       = {Vincent Conitzer and
                  Christian Kroer and
                  Debmalya Panigrahi and
                  Okke Schrijvers and
                  Nicol{\'{a}}s E. Stier Moses and
                  Eric Sodomka and
                  Christopher A. Wilkens},
  title        = {Pacing Equilibrium in First Price Auction Markets},
  journal      = {Management Science},
  volume       = {68},
  number       = {12},
  pages        = {8515--8535},
  year         = {2022},
  url          = {https://doi.org/10.1287/mnsc.2022.4310},
  doi          = {10.1287/MNSC.2022.4310},
  bibsource    = {dblp computer science bibliography, https://dblp.org}
}

@article{DBLP:journals/jacm/BadanidiyuruKS18,
  author       = {Ashwinkumar Badanidiyuru and
                  Robert Kleinberg and
                  Aleksandrs Slivkins},
  title        = {Bandits with Knapsacks},
  journal      = {Journal of the ACM (JACM)},
  volume       = {65},
  number       = {3},
  pages        = {13:1--13:55},
  year         = {2018},
  url          = {https://doi.org/10.1145/3164539},
  doi          = {10.1145/3164539},
  bibsource    = {dblp computer science bibliography, https://dblp.org}
}

@article{DBLP:journals/jacm/ImmorlicaSSS22,
  author       = {Nicole Immorlica and
                  Karthik Abinav Sankararaman and
                  Robert E. Schapire and
                  Aleksandrs Slivkins},
  title        = {Adversarial Bandits with Knapsacks},
  journal      = {Journal of the ACM (JACM)},
  volume       = {69},
  number       = {6},
  pages        = {40:1--40:47},
  year         = {2022},
  url          = {https://doi.org/10.1145/3557045},
  doi          = {10.1145/3557045},
  bibsource    = {dblp computer science bibliography, https://dblp.org}
}

@inproceedings{DBLP:conf/icalp/DobzinskiL14,
  author    = {Shahar Dobzinski and
               Renato {Paes Leme}},
  editor    = {Javier Esparza and
               Pierre Fraigniaud and
               Thore Husfeldt and
               Elias Koutsoupias},
  title     = {Efficiency Guarantees in Auctions with Budgets},
  booktitle = {Automata, Languages, and Programming - 41st International Colloquium,
               {ICALP} 2014, Copenhagen, Denmark, July 8-11, 2014, Proceedings, Part
               {I}},
  series    = {Lecture Notes in Computer Science},
  volume    = {8572},
  pages     = {392--404},
  publisher = {Springer},
  year      = {2014},
  url       = {https://doi.org/10.1007/978-3-662-43948-7\_33},
  doi       = {10.1007/978-3-662-43948-7\_33},
  bibsource = {dblp computer science bibliography, https://dblp.org}
}

@inproceedings{DBLP:conf/innovations/Babaioff0HIL21,
  author    = {Moshe Babaioff and
               Richard Cole and
               Jason D. Hartline and
               Nicole Immorlica and
               Brendan Lucier},
  editor    = {James R. Lee},
  title     = {Non-Quasi-Linear Agents in Quasi-Linear Mechanisms (Extended Abstract)},
  booktitle = {12th Innovations in Theoretical Computer Science Conference, {ITCS}
               2021, January 6-8, 2021, Virtual Conference},
  series    = {LIPIcs},
  volume    = {185},
  pages     = {84:1--84:1},
  publisher = {Schloss Dagstuhl - Leibniz-Zentrum f{\"{u}}r Informatik},
  year      = {2021},
  url       = {https://doi.org/10.4230/LIPIcs.ITCS.2021.84},
  doi       = {10.4230/LIPIcs.ITCS.2021.84},
  bibsource = {dblp computer science bibliography, https://dblp.org}
}

@article{DBLP:journals/mansci/BalseiroG19,
  author       = {Santiago R. Balseiro and
                  Yonatan Gur},
  title        = {Learning in Repeated Auctions with Budgets: Regret Minimization and Equilibrium},
  journal      = {Management Science},
  volume       = {65},
  number       = {9},
  pages        = {3952--3968},
  year         = {2019},
  url          = {https://doi.org/10.1287/mnsc.2018.3174},
  doi          = {10.1287/MNSC.2018.3174},
  bibsource    = {dblp computer science bibliography, https://dblp.org}
}

@article{DBLP:journals/mansci/BalseiroKK23,
  author       = {Santiago R. Balseiro and
                  Christian Kroer and
                  Rachitesh Kumar},
  title        = {Contextual Standard Auctions with Budgets: Revenue Equivalence and Efficiency Guarantees},
  journal      = {Management Science},
  volume       = {69},
  number       = {11},
  pages        = {6837--6854},
  year         = {2023},
  url          = {https://doi.org/10.1287/mnsc.2023.4719},
  doi          = {10.1287/MNSC.2023.4719},
  bibsource    = {dblp computer science bibliography, https://dblp.org}
}

@inproceedings{DBLP:conf/stoc/SyrgkanisT13,
  author    = {Vasilis Syrgkanis and
               {\'{E}}va Tardos},
  editor    = {Dan Boneh and
               Tim Roughgarden and
               Joan Feigenbaum},
  title     = {Composable and efficient mechanisms},
  booktitle = {Symposium on Theory of Computing Conference, STOC'13, Palo Alto, CA,
               USA, June 1-4, 2013},
  pages     = {211--220},
  publisher = {{ACM}},
  year      = {2013},
  url       = {https://doi.org/10.1145/2488608.2488635},
  doi       = {10.1145/2488608.2488635},
  bibsource = {dblp computer science bibliography, https://dblp.org}
}

@inproceedings{DBLP:conf/wine/AggarwalBM19,
  author    = {Gagan Aggarwal and
               Ashwinkumar Badanidiyuru and
               Aranyak Mehta},
  editor    = {Ioannis Caragiannis and
               Vahab S. Mirrokni and
               Evdokia Nikolova},
  title     = {Autobidding with Constraints},
  booktitle = {Web and Internet Economics - 15th International Conference, {WINE}
               2019, New York, NY, USA, December 10-12, 2019, Proceedings},
  series    = {Lecture Notes in Computer Science},
  volume    = {11920},
  pages     = {17--30},
  publisher = {Springer},
  year      = {2019},
  url       = {https://doi.org/10.1007/978-3-030-35389-6\_2},
  doi       = {10.1007/978-3-030-35389-6\_2},
  bibsource = {dblp computer science bibliography, https://dblp.org},
}

@article{DBLP:journals/ior/ConitzerKSM22,
  author       = {Vincent Conitzer and
                  Christian Kroer and
                  Eric Sodomka and
                  Nicol{\'{a}}s E. Stier Moses},
  title        = {Multiplicative Pacing Equilibria in Auction Markets},
  journal      = {Operations Research},
  volume       = {70},
  number       = {2},
  pages        = {963--989},
  year         = {2022},
  url          = {https://doi.org/10.1287/opre.2021.2167},
  doi          = {10.1287/OPRE.2021.2167},
  bibsource    = {dblp computer science bibliography, https://dblp.org}
}

@inproceedings{DBLP:conf/www/DengHLZ22,
  author    = {Xiaotie Deng and
               Xinyan Hu and
               Tao Lin and
               Weiqiang Zheng},
  editor    = {Fr{\'{e}}d{\'{e}}rique Laforest and
               Rapha{\"{e}}l Troncy and
               Elena Simperl and
               Deepak Agarwal and
               Aristides Gionis and
               Ivan Herman and
               Lionel M{\'{e}}dini},
  title     = {Nash Convergence of Mean-Based Learning Algorithms in First Price
               Auctions},
  booktitle = {{WWW} '22: The {ACM} Web Conference 2022, Virtual Event, Lyon, France,
               April 25 - 29, 2022},
  pages     = {141--150},
  publisher = {{ACM}},
  year      = {2022},
  url       = {https://doi.org/10.1145/3485447.3512059},
  doi       = {10.1145/3485447.3512059},
  bibsource = {dblp computer science bibliography, https://dblp.org}
}

@inproceedings{DBLP:conf/www/KolumbusN22,
  author    = {Yoav Kolumbus and
               Noam Nisan},
  editor    = {Fr{\'{e}}d{\'{e}}rique Laforest and
               Rapha{\"{e}}l Troncy and
               Elena Simperl and
               Deepak Agarwal and
               Aristides Gionis and
               Ivan Herman and
               Lionel M{\'{e}}dini},
  title     = {Auctions between Regret-Minimizing Agents},
  booktitle = {{WWW} '22: The {ACM} Web Conference 2022, Virtual Event, Lyon, France,
               April 25 - 29, 2022},
  pages     = {100--111},
  publisher = {{ACM}},
  year      = {2022},
  url       = {https://doi.org/10.1145/3485447.3512055},
  doi       = {10.1145/3485447.3512055},
  bibsource = {dblp computer science bibliography, https://dblp.org}
}

@article{DBLP:journals/corr/GaitondeLLLS22,
  author     = {Jason Gaitonde and
                Yingkai Li and
                Bar Light and
                Brendan Lucier and
                Aleksandrs Slivkins},
  title      = {Budget Pacing in Repeated Auctions: Regret and Efficiency without
                Convergence},
  journal    = {CoRR},
  volume     = {abs/2205.08674},
  year       = {2022},
  url        = {https://doi.org/10.48550/arXiv.2205.08674},
  doi        = {10.48550/arXiv.2205.08674},
  eprinttype = {arXiv},
  eprint     = {2205.08674},
  bibsource  = {dblp computer science bibliography, https://dblp.org}
}

@article{DBLP:journals/ftml/Slivkins19,
  author    = {Aleksandrs Slivkins},
  title     = {Introduction to Multi-Armed Bandits},
  journal   = {Foundations and Trends{\textregistered} in Machine Learning},
  volume    = {12},
  number    = {1-2},
  pages     = {1--286},
  year      = {2019},
  url       = {https://doi.org/10.1561/2200000068},
  doi       = {10.1561/2200000068},
  bibsource = {dblp computer science bibliography, https://dblp.org}
}

@article{DBLP:journals/ior/BalseiroKMM21,
  author    = {Santiago R. Balseiro and
               Anthony Kim and
               Mohammad Mahdian and
               Vahab S. Mirrokni},
  title     = {Budget-Management Strategies in Repeated Auctions},
  journal   = {Operations Research},
  volume    = {69},
  number    = {3},
  pages     = {859--876},
  year      = {2021},
  url       = {https://doi.org/10.1287/opre.2020.2073},
  doi       = {10.1287/opre.2020.2073},
  bibsource = {dblp computer science bibliography, https://dblp.org}
}

@article{despotakis2021first,
  title     = {First-price auctions in online display advertising},
  author    = {Despotakis, Stylianos and Ravi, R and Sayedi, Amin},
  journal   = {Journal of Marketing Research},
  volume    = {58},
  number    = {5},
  pages     = {888--907},
  year      = {2021},
  publisher = {SAGE Publications Sage CA: Los Angeles, CA}
}

@misc{wong_2021,
  title     = {Moving AdSense to a first-price auction},
  url       = {https://blog.google/products/adsense/our-move-to-a-first-price-auction/},
  journal   = {Google},
  publisher = {Google},
  author    = {Wong, Matt},
  year      = {2021},
  month     = {10}
}

@misc{yuen_2022,
  title   = {Programmatic advertising in 2022: Digital Display Ads Industry},
  url     = {https://www.insiderintelligence.com/insights/programmatic-digital-display-ad-spending},
  journal = {Insider Intelligence},
  author  = {Yuen, Meaghan},
  year    = {2022},
  month   = {5}
}

@inproceedings{DBLP:conf/aaai/0004GLMS21,
  author    = {Zhe Feng and
               Guru Guruganesh and
               Christopher Liaw and
               Aranyak Mehta and
               Abhishek Sethi},
  title     = {Convergence Analysis of No-Regret Bidding Algorithms in Repeated Auctions},
  booktitle = {Thirty-Fifth {AAAI} Conference on Artificial Intelligence, {AAAI}
               2021, Thirty-Third Conference on Innovative Applications of Artificial
               Intelligence, {IAAI} 2021, The Eleventh Symposium on Educational Advances
               in Artificial Intelligence, {EAAI} 2021, Virtual Event, February 2-9,
               2021},
  pages     = {5399--5406},
  publisher = {{AAAI} Press},
  year      = {2021},
  url       = {https://ojs.aaai.org/index.php/AAAI/article/view/16680},
  bibsource = {dblp computer science bibliography, https://dblp.org}
}

@inproceedings{DBLP:conf/icml/CastiglioniCK22,
  author    = {Matteo Castiglioni and
               Andrea Celli and
               Christian Kroer},
  editor    = {Kamalika Chaudhuri and
               Stefanie Jegelka and
               Le Song and
               Csaba Szepesv{\'{a}}ri and
               Gang Niu and
               Sivan Sabato},
  title     = {Online Learning with Knapsacks: the Best of Both Worlds},
  booktitle = {International Conference on Machine Learning, {ICML} 2022, 17-23 July
               2022, Baltimore, Maryland, {USA}},
  series    = {Proceedings of Machine Learning Research},
  volume    = {162},
  pages     = {2767--2783},
  publisher = {{PMLR}},
  year      = {2022},
  url       = {https://proceedings.mlr.press/v162/castiglioni22a.html},
  bibsource = {dblp computer science bibliography, https://dblp.org}
}

@article{DBLP:journals/corr/abs-2207-04690,
  author    = {Zhaohua Chen and
               Chang Wang and
               Qian Wang and
               Yuqi Pan and
               Zhuming Shi and
               Chuyue Tang and
               Zheng Cai and
               Yukun Ren and
               Zhihua Zhu and
               Xiaotie Deng},
  title     = {Dynamic Budget Throttling in Repeated Second-Price Auctions},
  journal   = {CoRR},
  volume    = {abs/2207.04690},
  year      = {2022},
  url       = {https://doi.org/10.48550/arXiv.2207.04690},
  doi       = {10.48550/arXiv.2207.04690},
  eprinttype = {arXiv},
  eprint    = {2207.04690},
  bibsource = {dblp computer science bibliography, https://dblp.org}
}

@article{DBLP:journals/corr/abs-2203-16816,
  author    = {Zhaohua Chen and
               Xiaotie Deng and
               Jicheng Li and
               Chang Wang and
               Mingwei Yang},
  title     = {Budget-Constrained Auctions with Unassured Priors},
  journal   = {CoRR},
  volume    = {abs/2203.16816},
  year      = {2022},
  url       = {https://doi.org/10.48550/arXiv.2203.16816},
  doi       = {10.48550/arXiv.2203.16816},
  eprinttype = {arXiv},
  eprint    = {2203.16816},
  bibsource = {dblp computer science bibliography, https://dblp.org}
}

@article{DBLP:journals/corr/LucierPSZ23,
  author       = {Brendan Lucier and
                  Sarath Pattathil and
                  Aleksandrs Slivkins and
                  Mengxiao Zhang},
  title        = {Autobidders with Budget and {ROI} Constraints: Efficiency, Regret,
                  and Pacing Dynamics},
  journal      = {CoRR},
  volume       = {abs/2301.13306},
  year         = {2023},
  url          = {https://doi.org/10.48550/arXiv.2301.13306},
  doi          = {10.48550/arXiv.2301.13306},
  eprinttype    = {arXiv},
  eprint       = {2301.13306},
  bibsource    = {dblp computer science bibliography, https://dblp.org}
}

@inproceedings{DBLP:conf/www/DengMMZ21,
  author       = {Yuan Deng and
                  Jieming Mao and
                  Vahab S. Mirrokni and
                  Song Zuo},
  editor       = {Jure Leskovec and
                  Marko Grobelnik and
                  Marc Najork and
                  Jie Tang and
                  Leila Zia},
  title        = {Towards Efficient Auctions in an Auto-bidding World},
  booktitle    = {{WWW} '21: The Web Conference 2021, Virtual Event / Ljubljana, Slovenia,
                  April 19-23, 2021},
  pages        = {3965--3973},
  publisher    = {{ACM} / {IW3C2}},
  year         = {2021},
  url          = {https://doi.org/10.1145/3442381.3450052},
  doi          = {10.1145/3442381.3450052},
  bibsource    = {dblp computer science bibliography, https://dblp.org}
}

@inproceedings{DBLP:conf/colt/FikiorisT23,
  author       = {Giannis Fikioris and
                  {\'{E}}va Tardos},
  editor       = {Gergely Neu and
                  Lorenzo Rosasco},
  title        = {Approximately Stationary Bandits with Knapsacks},
  booktitle    = {The Thirty Sixth Annual Conference on Learning Theory, {COLT} 2023,
                  12-15 July 2023, Bangalore, India},
  series       = {Proceedings of Machine Learning Research},
  volume       = {195},
  pages        = {3758--3782},
  publisher    = {{PMLR}},
  year         = {2023},
  url          = {https://proceedings.mlr.press/v195/fikioris23a.html},
  bibsource    = {dblp computer science bibliography, https://dblp.org}
}

@inproceedings{DBLP:conf/soda/LemeST12,
  author       = {Renato Paes Leme and
                  Vasilis Syrgkanis and
                  {\'{E}}va Tardos},
  editor       = {Yuval Rabani},
  title        = {Sequential auctions and externalities},
  booktitle    = {Proceedings of the Twenty-Third Annual {ACM-SIAM} Symposium on Discrete
                  Algorithms, {SODA} 2012, Kyoto, Japan, January 17-19, 2012},
  pages        = {869--886},
  publisher    = {{SIAM}},
  year         = {2012},
  url          = {https://doi.org/10.1137/1.9781611973099.70},
  doi          = {10.1137/1.9781611973099.70},
  bibsource    = {dblp computer science bibliography, https://dblp.org}
}

@article{DBLP:journals/geb/FeldmanFGL20,
  author       = {Michal Feldman and
                  Hu Fu and
                  Nick Gravin and
                  Brendan Lucier},
  title        = {Simultaneous auctions without complements are (almost) efficient},
  journal      = {Games Econ. Behav.},
  volume       = {123},
  pages        = {327--341},
  year         = {2020},
  url          = {https://doi.org/10.1016/j.geb.2015.11.009},
  doi          = {10.1016/J.GEB.2015.11.009},
  bibsource    = {dblp computer science bibliography, https://dblp.org}
}

\appendix{}
\section{Calculations for the proof of Theorem \ref{thm:guar}}
\label{sec:app:calc}

We first show that $\l_1$ and $\l_2$ are in the range $(0,1)$. The first is obvious, since $\l_1 = \frac{1}{\sqrt{\g + 1}}$. 
To prove that $\l_2 = \g W\left( - e^{-2/\g}/\g \right) + 1 \in (0,1)$, we first note that $W(x)$ is an increasing function in $x$ and that $W(x) \in (-1,0)$ when $x \in (-1/e,0)$. This immediately proves that $\l_2 < 1$. We now notice that
\begin{align*}
    W\left( - e^{-2/\g}/\g \right)
    >
    W\left( - e^{-1/\g}/\g \right)
    =
    - \frac{1}{\g}
\end{align*}
where the first inequality follows from the fact that $W$ is strictly increasing and the second from the fact that $W(xe^x) = x$, by definition for $x \ge -1$. This proves that $\l_2 > 0$.

Now let $\g_0$ be the solution to
\begin{align*}
    1 + \frac{1}{\l_1} = \frac{\g}{c(\l_1) \l_1}
    \;\Longleftrightarrow\;
    1 + \sqrt{\g + 1} = 
    - \g \sqrt{\g + 1} \ln\left( 1 - \frac{1}{\sqrt{\g + 1}} \right)
\end{align*}

We prove that $\g_0$ is the unique solution by noticing that $1 + \sqrt{\g + 1} + \g \sqrt{\g + 1} \ln\left( 1 - \frac{1}{\sqrt{\g + 1}} \right)$ is strictly decreasing for $\g \ge 1$ (can be proved in Mathematica). Numerically we can solve the above equation to see that $\g_0 \approx 1.73$. These observations also prove that for $\g < \g_0$
\begin{align*}
    1 + \frac{1}{\l_1}
    >
    \frac{\g}{c(\l_1)\l_1}
\end{align*}
as needed by the proof in \cref{thm:guar}.

We now prove that for $\g \ge \g_0$ it holds that
\begin{align}\label{eq:app:1}
    1 + \frac{1}{\l_2}
    \ge
    \frac{\g\l_2}{1-\l_2}
\end{align}

First, we notice that the equality holds when $\g = \g_0$, since from before this was the unique solution to the system $1 + \frac{1}{\l} = \frac{\g\l}{1-\l} = \frac{\g}{c(\l)\l}$. To prove the inequality all we need to prove is that $1 + \frac{1}{\l_2}$ is increasing in $\g$ and $\frac{\g\l_2}{1-\l_2}$ is decreasing in $\g$. We do both of these by proving that $\l_2$ is decreasing in $\g$. The derivative of $\l_2$ (calculated in Mathematica) is
\begin{align*}
    \frac{d \l_2}{d\g}
    =
    \frac{W\left(-\frac{e^{-2/\gamma }}{\gamma }\right)}{\g}
    \frac{2 + \g W\left(-\frac{e^{-2/\gamma }}{\gamma }\right)}{1 + W\left(-\frac{e^{-2/\gamma }}{\gamma }\right)}.
\end{align*}

The above quantity is strictly negative, because of the observations made at the beginning of the section:
\begin{itemize}
    \item $W\left(-\frac{e^{-2/\gamma }}{\gamma }\right) \in (-1, 0)$.
    \item $2 + \g W\left(-\frac{e^{-2/\gamma }}{\gamma }\right) = 1 + \l_2 \in (1, 2)$.
\end{itemize}

These two observations prove \eqref{eq:app:1}.

We now point out that $1 + 1/\l_2 = \g + 1/2 + O(1/\g)$. One can verify that this is the case using Mathematica.

Finally, we need to bound the factor in front of the regret term for $\l \in \{\l_1, \l_2\}$,
\begin{align*}
    \max\left\{\frac{\l}{1-\l} , \frac{1}{c(\l)\l}\right\}
    +
    \frac{1}{\g\l}
\end{align*}

We use the fact proven above, that for $\l \in \{\l_1, \l_2\}$, 
\begin{align*}
    1 + \frac{1}{\l}
    =
    \max\left\{
        1 + \frac{1}{\l}, \frac{\g\l}{1-\l}, \frac{\g}{c(\l)\l}
    \right\}
\end{align*}
to get that
\begin{align*}
    \max\left\{\frac{\l}{1-\l} , \frac{1}{c(\l)\l}\right\}
    +
    \frac{1}{\g\l}
    \le
    \frac{1}{\g} + \frac{1}{\g\l}
    +
    \frac{1}{\g\l}
    =
    O(1)
\end{align*}
where the final equality holds because either $\l = \l_1 = \Theta(1/\sqrt \g)$ or for $\l = \l_2 = \Theta(1/\g)$





\end{document}